\documentclass[lettersize,journal]{IEEEtran}
\usepackage{setspace}
\usepackage{amsmath,amsfonts}
\usepackage{array}

\usepackage{textcomp}
\usepackage{stfloats}
\usepackage{url}
\usepackage{verbatim}
\usepackage{graphicx}
\def\BibTeX{{\rm B\kern-.05em{\sc i\kern-.025em b}\kern-.08em
		T\kern-.1667em\lower.7ex\hbox{E}\kern-.125emX}}

\usepackage{algorithm}
\usepackage{array}
\usepackage{algorithm}
\usepackage{amsmath}
\usepackage{bbm}

\usepackage{amsmath}
\usepackage{pdfpages}
\usepackage{graphicx}
\usepackage{epstopdf}
\usepackage{stackengine,scalerel}
\usepackage{amsmath,amsthm,amssymb}
\usepackage{color,soul}
\usepackage{amsfonts}
\usepackage{epsfig}
\usepackage{amssymb}
\usepackage{cite}
\usepackage{subfigure}
\usepackage{multirow}
\usepackage{rotating}
\usepackage{tabularx}
\usepackage{array}
\usepackage{amsfonts}
\usepackage{epsfig}
\usepackage[all]{xy}
\usepackage{tabularx}
\usepackage{amsmath}
\usepackage{amsfonts}
\usepackage{epstopdf}
\usepackage{amssymb}
\usepackage{cite}
\usepackage{subfigure}
\usepackage{multirow}
\usepackage{rotating}
\usepackage{tabularx}
\usepackage{array}
\usepackage{setspace}
\usepackage{morefloats}
\usepackage{float}
\usepackage[Euler]{upgreek}
\usepackage{mathtools}
\usepackage[mathscr]{eucal}

\makeatletter
\newcommand*{\rom}[1]{\expandafter\@slowromancap\romannumeral #1@}
\makeatother
\usepackage{amsmath,amssymb}
\usepackage{amsmath}
\usepackage{algorithm}
\usepackage{algpseudocode}
\usepackage{float}
\usepackage{cite}
\usepackage{algorithm,algpseudocode}
\usepackage{caption}
\makeatletter
\newcommand{\algmargin}{\the\ALG@thistlm}
\makeatother
\newlength{\whilewidth}
\algdef{SE}[parWHILE]{parWhile}{EndparWhile}[1]
{\parbox[t]{\dimexpr\linewidth-\algmargin}{%
		\hangindent\whilewidth\strut\algorithmicwhile\ #1\ \algorithmicdo\strut}}{\algorithmicend\ \algorithmicwhile}%
\algnewcommand{\parState}[1]{\State%
	\parbox[t]{\dimexpr\linewidth-\algmargin}{\strut #1\strut}}
\makeatletter
\def \BState{\State \hskip - \ALG@thistlm}
\makeatother
\newtheorem{definition}{Definition}
\newtheorem{remark}{Remark}
\begin{document}
	\title{Precise HDV Positioning through Safety-Aware Integrated Sensing and Communication in a Value-of-Information-Driven 6G V2X System 
		%		Resource-Efficient Multi-Platooning Control of Connected
		%		Automated Vehicles Based on AoI Over Multi-Lane Freeways
\author{Mohammad Reza~Abedi, \IEEEmembership{Member}, IEEE, Zahra~Rashidi, and Nader~Mokari, \IEEEmembership{Senior Member}, IEEE,
	Hamid~Saeedi, \IEEEmembership{Member}, IEEE, and
	Nizar Zorba,~\IEEEmembership{Senior Member,~IEEE} 
}
\thanks{Mohammadreza Abedi, Zahra Rashidi, Nader Mokari  are with the Department of Electrical Engineering, Tarbiat Modares	University, Tehran, Iran (e-mail: Mohammadreza\_abedi@modares.ac.ir, zrashi@modares.ac.ir, and nader.mokari@modares.ac.ir). Hamid Saeedi is with College of Engineering and Technology, University of Doha for Science and Technology, Qatar (e-mail: hamid.saeedi@udst.edu.qa). Nizar Zorba is with the College of Engineering, Qatar University, Doha, Qatar (e-mail: nizarz@qu.edu.qa).}
}
	\maketitle
	%\thanks{Manuscript received XXX, XX, 2020; revised XXX, XX, 2020.}}
\markboth{IEEE Transactions on Wireless Communications}%
%{Submitted paper}
%\markboth{IEEE Journal on Selected Areas in Communication,~Vol.~XX, No.~XX, XXX~2020}
{Submitted paper}
\maketitle
\makeatletter
\def\endthebibliography{%
	\def\@noitemerr{\@latex@warning{Empty `thebibliography' environment}}%
	\endlist
}
\makeatother
	\begin{abstract}
Recent advancements in Integrated Sensing and Communications (ISAC) have unlocked new potential for addressing the dual demands of high-resolution positioning and reliable communication in 6G Vehicle-to-Everything (V2X) networks. These capabilities are vital for transmitting safety-critical data from Connected Autonomous Vehicles (CAVs) to improve metrics such as Time to Collision (TTC) and reduce the Collision Risk (CR) ratio. However, limited radio resources and interference remain major obstacles to achieving both precision and capacity simultaneously. The challenge intensifies in mixed-traffic scenarios involving Human-Driven Vehicles (HDVs), which lack connectivity and cannot share their status or positioning. Additionally, CAV sensors are limited in range and accuracy, making detection of HDVs unreliable. ISAC plays a pivotal role here by enabling the sensing of HDV positions via shared communication infrastructure, improving environmental awareness. To address these challenges, this paper proposes a novel Value of Information (VoI) metric that prioritizes the transmission of safety-critical data. The joint sensing-communication-control problem is modeled as a two-time-scale sequential decision process and solved using a Multi-Agent Distributed Deterministic Policy Gradient (MADDPG) algorithm. By focusing on high-VoI data, the framework reduces complexity and optimizes network and traffic resource usage. Simulations show that the proposed approach significantly reduces the CR ratio by at least 33\% and improves the TTC by up to 66\%, demonstrating its effectiveness in enhancing safety and efficiency in mixed-autonomy environments.
\end{abstract}

\section{Introduction}
\subsection{Motivations and State of the Art}
%The rapid advancement of vehicle automation offers transformative potential in improving efficiency, safety, and driving comfort. Among the key components of automated driving systems, longitudinal control mechanisms such as Adaptive Cruise Control (ACC) have become central to Automated Vehicle (AV) research. These systems address critical challenges, including enhancing safety, ensuring string stability, adapting to individual driving preferences, and mitigating response delays. ACC technology, now widely available to the public \cite{10531066}, extends beyond individual vehicle control to encompass broader applications in traffic management. As mobile actuators, AVs have demonstrated their ability to improve large-scale traffic flow and stability, reduce congestion, and minimize emissions and noise pollution, as shown in experimental studies \cite{10806635,10769014}. These multiple impacts underscore AVs' pivotal role in enhancing individual driving experiences and optimizing overall traffic dynamics. 
The rapid growth of vehicle automation highlights the importance of longitudinal control systems like Adaptive Cruise Control (ACC). ACC not only improves safety, comfort, and string stability for individual vehicles but also benefits large-scale traffic by enhancing flow, reducing congestion, and minimizing emissions. These impacts demonstrate the key role of Automated Vehicles (AVs) in advancing both driving experiences and overall traffic efficiency. Cooperative Adaptive Cruise Control (CACC) surpasses ACC by utilizing Vehicle-to-Infrastructure (V2I) and Vehicle-to-Vehicle (V2V) communication to exchange key data (e.g., position, speed, acceleration), enabling coordinated and efficient driving. Within the Vehicle-to-Everything (V2X) framework, AVs fuse external information with onboard sensor data, thereby enhancing vehicle guidance and significantly improving traffic safety.
%While ACC systems have proven effective, Cooperative Adaptive Cruise Control (CACC) offers significantly superior performance. By leveraging Vehicle-to-Infrastructure (V2I) and Vehicle-to-Vehicle (V2V) communication networks, CACC enables vehicles to share critical data—such as position, speed, and acceleration—resulting in more efficient and coordinated driving. Through Vehicle-to-Everything (V2X) framework, AVs can integrate this exogenous information with their locally obtained data from onboard sensors. This fusion of external and internal data not only enhances vehicle guidance but also significantly improves overall traffic safety.  
%Accurate vehicle positioning is fundamental to AV safety, particularly in environments where Global Positioning System (GPS) functionality is impaired, such as tunnels or underground parking facilities. While GPS remains the prevalent solution for vehicle positioning, it is not always reliable \cite{8930502}. Furthermore, mixed traffic conditions—where autonomous and human-driven vehicles share the road—add complexity. Seamless interaction between these vehicle types necessitates an understanding of human driving behavior \cite{10478278}. To address these challenges, emerging mobile communication systems such as 6th Generation (6G) networks are increasingly integrating sensing and communication capabilities.  
Accurate positioning is crucial for AV safety, especially where GPS is unreliable (e.g., tunnels, parking). Mixed traffic with human-driven vehicles adds complexity, requiring behavioral understanding. To overcome these issues, emerging 6G networks integrate sensing and communication for enhanced positioning and safety.

Integrated Sensing and Communications (ISAC) is a transformative paradigm for vehicular networks that enables object detection without extra spectrum cost, aligning with 6G’s IMT-2030 vision \cite{10639496,10845207,10556618,9791349,9728752,9724187,10960374,10032141,10158711}. ISAC improves positioning when GPS is weak or unavailable, supports localization of GPS-lacking vehicles, and enables V2X information sharing—enhancing situational awareness, coordination, and road safety. Recent advances such as mmWave (wide spectrum, high data rates, fine sensing resolution) and massive Multiple-Input Multiple-Output (mMIMO) (directional beamforming, spatial multiplexing) mitigate mmWave path loss while boosting positioning accuracy. Their synergy makes simultaneous data transmission and sensing feasible, positioning ISAC as a key enabler of future intelligent transportation systems.

Although advanced technologies improve vehicular positioning and communication, practical deployment is limited by bandwidth constraints and the infeasibility of transmitting all vehicles’ data in dense traffic. To address this, we introduce the Value of Information (VoI) metric, which prioritizes data based on its contribution to safety and situational awareness, unlike the Age of Information (AoI) that only considers freshness. This paper investigates which vehicles’ positioning data should be prioritized for radar estimation and transmission under resource constraints: \textit{Which vehicles’ positioning information should be prioritized for radar-based estimation and transmission to maximize safety and situational awareness under constrained radio resources?} We propose a unified VoI-driven framework that enables selective radar sensing and V2I transmission of high-value information, while constructing a compact state space for Deep Reinforcement Learning (DRL)-based control, ensuring efficient resource allocation and real-time decision-making. Therefore, we pose a secondary question: \textit{How should transport and radio network parameters be selected to construct a compact yet informative state space for DRL-based autonomous vehicle control?}
\subsection{Related Works}
V2I networks play a crucial role in 6G cellular systems, particularly with the growing ISAC. Roadside Base Stations (BSs) are expected to provide both data transmission and vehicle tracking capabilities. Existing research primarily focuses on two sensing paradigms—active and passive sensing—within V2I systems \cite{9171304}. Considerable efforts have been dedicated to the development of integrated active sensing and communication mechanisms, where full-duplex BSs simultaneously transmit downlink (DL) data and monitor vehicle motion via echo signals \cite{9171304,9246715,9724174}. For instance, \cite{9171304} proposed a power allocation scheme to minimize the Cramér-Rao Lower Bound (CRLB) under a sum-rate constraint in multi-vehicle scenarios. Expanding on this, \cite{9724174} tackled multi-user interference challenges by introducing a Weighted sum Mean Square Error minimization (WMMSE)-based iterative algorithm for optimized DL beamforming.

While ISAC-based V2I scenarios have received significant attention, several critical challenges hinder their practical implementation. A major limitation in prior works is the predominant focus on single-Roadside Units (RSU) V2I scenarios \cite{9791349},\cite{9171304},\cite{10061429}. Given that vehicles are continuously mobile, ensuring seamless sensing and communication services requires a multi-RSU deployment strategy. However, most existing solutions fail to effectively address the coordination among multiple RSUs, leading to service disruptions.

Some studies have explored multi-RSU V2I scenarios under conventional radar-sensing architectures. However, they typically emphasize sensing accuracy while neglecting communication performance \cite{9229189,9800700}. For instance, \cite{9229189} proposed a multi-sensor, multi-vehicle positioning and tracking framework for autonomous driving, showing that cooperation among RSUs enhances positioning accuracy. Similarly, \cite{9800700} developed a multi-RSU collaborative radar sensing network leveraging signal-level fusion technology to assist vehicles in perceiving their surroundings. Despite these advancements, these methods overlook the challenges of radio resource allocation and efficient beamforming for real-time communication and tracking in dynamic traffic environments.

Recent works have proposed innovative ISAC-based frameworks to overcome some of these challenges. For instance, \cite{10504613} introduced a Nonlinear Self-Interference Cancellation (NSIC) scheme for ISAC-assisted V2X networks, considering beam tracking. While effective in reconstructing self-interference using CSI-based path projection, this approach fails to account for practical issues such as CSI estimation errors, processing delays, and external interference. Another study \cite{9820762} proposed a multi-vehicle tracking and ID association scheme using ISAC signals, where RSUs transmit multi-beam signals to estimate vehicle positions and velocities while employing Kullback-Leibler Divergence (KLD) for ID association. However, this approach assumes an idealized environment and does not fully address real-world factors such as environmental noise, sudden vehicle maneuvers, and signal distortions.

Further, \cite{10677488} developed a multi-beam object positioning framework to enhance sensing performance in mmWave MIMO ISAC systems for connected autonomous vehicles, optimizing the sensing beampattern gain under SINR, power, and hardware constraints. However, the scheme primarily focuses on improving sensing rather than addressing communication and radio resource management. Similarly, \cite{10433790} proposed an ISAC-assisted collision avoidance mechanism that utilizes mmWave MIMO and beamforming for simultaneous communication and environmental sensing. However, its power allocation scheme does not consider the VoI in prioritizing safety-critical data transmission. Additionally, \cite{10502156} introduced ISAC-assisted frame structures for NR-V2X communications, reducing pilot overhead while improving beam management, yet the work lacks an adaptive mechanism for dynamically allocating radio resources in real-time scenarios. Lastly, \cite{10738493} proposed an ISAC-based beam tracking scheme for multi-RSU V2I systems, integrating an unscented Kalman filter and CoMP-based resource allocation to minimize inter-region interference. However, its approach does not explicitly address the optimization of radio resources based on the importance of transmitted information.
\subsection{Contributions}
The innovations of the paper, presented in a technical and precise manner, are as follows:
\begin{itemize}
\item \textit{ISAC-Assisted Position Estimation for Human-Driven Vehicles}: Unlike prior studies focusing on autonomous vehicle tracking, our approach employs ISAC technology to accurately estimate the position of human-driven vehicles and relay this information via V2I links. This innovation significantly improves the coordination and safety of mixed traffic environments, particularly in GPS-denied scenarios.
\item \textit{Integration of VoI in ITS Systems}: Unlike conventional methods that treat all transmitted data equally, our work introduces the concept of VoI in ITSs. By distinguishing and prioritizing high-value information that enhances safety, our approach ensures efficient allocation of communication resources, leading to improved reliability and system performance.
\item \textit{Radio Resource Management for ISAC-Based V2I Networks}: Addressing the challenge of limited radio resources, we propose a novel beamforming and resource allocation mechanism that prioritizes the transmission of critical safety information. Unlike existing schemes, which allocate resources based on sensing accuracy alone, our method dynamically optimizes beamforming parameters while ensuring that the most valuable data is transmitted efficiently.
\end{itemize}
%By tackling these challenges, our proposed framework enhances both sensing and communication performance in ISAC-assisted V2X networks, ensuring robust operation in dynamic vehicular environments.

%\subsection{Paper Organization}
%The structure of this paper is as follows: Section \ref{System_Model} provides a detailed description of the system model. In Section \ref{Value-Info}, we introduce a framework for quantifying the value of exogenous information. Sections \ref{Problem Formulation} and \ref{Solution} outline the problem formulation and present the solution methodology based on the proposed MA-DDPG approach, respectively. The computational complexity analysis of the proposed solution is discussed in Section \ref{Signaling Overhead}. The performance evaluation results are presented in Section \ref{Simulation}. Finally, Section \ref{Conclusion} concludes the paper by summarizing the key findings.
The structure of this paper is as follows: Section \ref{System_Model} details the system model, while Section \ref{Value-Info} introduces a framework for quantifying the value of exogenous information. Section \ref{Problem Formulation} formulates the problem, and Section \ref{Solution} presents the MA-DDPG-based solution methodology. Section \ref{Signaling Overhead} discusses the computational complexity, followed by performance evaluation results in Section \ref{Simulation}. Finally, Section \ref{Conclusion} summarizes the key findings.

\section{System Model}\label{System_Model}
%\subsection{Basic Assumptions}
In an urban environment, a square region is equipped with a Cell-Free massive MIMO (CFmMIMO) system, which consists of $R$ RSUs strategically positioned along both sides of the road. Each RSU is outfitted with a massive Uniform Linear Array (ULA) consisting of $M_t$ transmit antennas and $M_r$ receive antennas, where $M_t=M_r=M=P\times Q$, and $M_t R \gg V$, where the symbol $\gg$ denotes that one quantity is much greater than another. The set of $V$ CAVs is denoted by $\mathcal{V}=\{v_1, v_2, \dots, v_V\}$, where $|\mathcal{V}|=V$, and each vehicle is indexed by $v$, where the notation $|.|$ represents the cardinality (size) of set. RSUs are connected to a Central Processing Unit (CPU) and support a Dual-function Radar Communication (DFRC) system operating in the millimeter-Wave (mmWave) band. This system simultaneously performs sensing and communication tasks, serving $V$ CAVs, each equipped with a single antenna. The set of RSUs is denoted by $\mathcal{R}=\{r_1, r_2, \dots, r_R\}$, with cardinality $|\mathcal{R}|=R$, and each RSU is indexed by $r$. A selected subset of these RSUs is used to serve each CAV via coherent joint transmissions within the user-centric cluster. The subset of RSUs used to serve vehicle $v$ is denoted by $\mathcal{R}_v$, where $|\mathcal{R}_v| = R_v$. RSUs are chosen based on their large-scale fading coefficients, which are the highest for vehicle $v$. The set of CAVs served by RSU $r$ is defined as $\mathcal{V}_r$, where $|\mathcal{V}_r|=V_r \leq M_t$. The road consists of $L$ lanes, each with a width of $D_l$ and a center position given by $y_l = D_l / 2 + (l-1)D_l$. The set of lanes is represented by $\mathcal{L} = \{l_1, l_2, \dots, l_L\}$, indexed by $l$, with $|\mathcal{L}| = L$.

To address the different time scales of transportation and radio network control parameters, a two-time scale control scheme is adopted to reduce signal processing complexity and overhead. The time axis is divided into several large-term time slots, each of duration $\Delta$, indexed by $\tau$. Each large-term time slot is further divided into $T$ short-term time slots, each with a duration of $\delta$, indexed by $t_\tau$. Based on the studies in \cite{9947033,10061429}, it is reasonable to assume that the state parameters of vehicles remain constant over a short time duration. Each CAV $v$ has a position $\textbf{q}^{l}_{v,t_\tau}=\left(x^{l}_{v,t_\tau}, y^{l}_{v,t_\tau}\right)$, where $x^{l}_{v,t_\tau}$ and $y^{l}_{v,t_\tau}$ represent the longitudinal and lateral positions during short time slot $t_\tau$, respectively. The velocity and acceleration of CAV $v$ on lane $l$ at short-term time $t_\tau$ are denoted by $\vartheta^{l}_{v,t_\tau}$ and $z^{l}_{v,t_\tau}$, respectively. The dynamic model of CAVs, based on their trajectories, is described as follows:
\begin{align}\label{eq-M-1}
	 &\dot{x}^l_{v,t_\tau}={\vartheta}^l_{v,t_\tau}\cos(\theta^l_{v,t_\tau}),~  \dot{y}^l_{v,t_\tau}={\vartheta}^l_{v,t_\tau}\sin(\theta^l_{v,t_\tau}),\\&\label{eq-M-3} \dot{\vartheta}^l_{v,t_\tau}=z^l_{v,t_\tau},~ \dot{\theta}^l_{v,t_\tau}=\tan(\alpha^l_{v,\tau})\vartheta^l_{v,t_\tau}/D_v,\\&\label{eq-M-5} \dot{z}^l_{v,t_\tau}=\frac{1}{\xi_v}z^l_{v,t_\tau}+\frac{1}{\xi_v}u^l_{v,\tau}, 
\end{align}
where $\theta^{l}_{v,t_\tau} \in [-\pi,\pi]$ denotes the heading angle of CAV $v$ in lane $l$ at short-term time $t_\tau$. The variables $\alpha^l_{v,\tau}$ and $u^l_{v,\tau}$ represent the control inputs, including the steering angle and acceleration (e.g., thrust or braking), respectively, which are designed at the large-term time scale $\tau$.
%	, as shown in Fig.\ref{Fig-wheel}. 
%	\begin{figure}[!t]
	%		\centering
	%		\includegraphics[width=1.7in]{Wheel.png}%
	%		\caption{Motion parameters of CAV $v$.}
	%		\label{Fig-wheel}
	%	\end{figure}
The parameters $\xi_v$ and $D_v$ represent the response time to a given set of control inputs and the physical length of CAV $v$, respectively. It is important to note that the same equations can also be applied to the preceding vehicle (predecessor) $(v-1)$ by substituting the subscript $v$ with $(v-1)$. To ensure safety—i.e., avoiding collisions with surrounding vehicles within a specific lane—and to satisfy the physical limitations of the vehicles, the following boundary constraints are imposed on the control variables of each CAV $v$ in lane $l$ over both long-term $\tau$ and short-term $t_\tau$ time horizons:
\begin{align}\label{eq-1}
	&\text{}~\Delta q^{l}_{v,t_\tau}\geq d^{l}_{v,t_\tau},
	\text{}~\vartheta^{l,\text{min}}\leq \vartheta^l_{v,t_\tau}\cos(\theta^l_{v,t_\tau}) \leq \vartheta^{l,\text{max}},\\\label{eq-3}&\text{}~-z^{\text{max}}\leq z^l_{v,t_\tau}\cos(\theta^l_{v,t_\tau})\leq z^{\text{max}},\text{}~-u^{\text{max}}\leq u^l_{v,\tau}\leq u^{\text{max}},\\\label{eq-5}&\text{}~-\alpha^{\text{max}}\leq \alpha^l_{v,\tau}\leq \alpha^{\text{max}},
\end{align}
where $\Delta q^{l}_{v,t_\tau} = \sqrt{(x^l_{v,t_\tau} - x^{l}_{(v-1),t_\tau})^2 + (y^l_{v,t_\tau} - y^{l}_{(v-1),t_\tau})^2}\\-D_v$ denotes the distance between CAV $v$ and the preceding CAV $(v-1)$ at lane $l$ at short-term time $t_{\tau}$, $d^{l}_{v,t_\tau} = d_0 + T_v (\vartheta^l_{v,t_\tau} - \vartheta^{l}_{(v-1),t_\tau})$ is the desired (safe) gap (i.e., the minimum distance to avoid a collision) between CAV $v$ and the preceding CAV $(v-1)$ at lane $l$ at short-term time $t_{\tau}$, $d_0$ is a constant standstill distance for each CAV, $T_v$ is the time gap for CAV $v$, $\vartheta^{l,\text{min}}$ ($\vartheta^{l,\text{max}}$) is the minimum (maximum) velocity at lane $l$, and $u^{\text{max}}$ and $\alpha^{\text{max}}$ are the maximum values of the control inputs, including acceleration and steering angle, respectively. The constraints (\ref{eq-1})-(\ref{eq-5}) describe the lower limit of the gap between adjacent CAVs, the range of distance, velocity, acceleration, and heading angle of the CAVs, respectively.

The control errors of CAV $v$, including spacing and velocity errors in lane $l$ at short-term time $t_\tau$, are defined as follows:
\begin{align}
	&e^{l}_{v,t_\tau}=\Delta q^{l}_{v,t_\tau}-d^{l}_{v,t_\tau}, \tilde{e}^{l}_{v,t_\tau}=\vartheta^l_{v,t_\tau}-\vartheta^{l}_{(v-1),t_\tau}.
\end{align}
%where $|.|$ denotes the absolute value.

Minimizing spacing and velocity errors is crucial to ensuring proper safety for all CAVs. However, it is important to note that only neighboring CAVs on the same lane significantly impact safety and should be considered. Therefore, we define a set of neighboring CAVs for each CAV $v$ on lane $l$ as $\mathcal{V}^l_{v,t_\tau} = {v^l_{v'} : \Delta q^{l}_{vv',t_\tau} \leq \Delta q^{\text{th}}}$, where $\Delta q^{l}_{vv',t_\tau} = \sqrt{(x^l_{v,t_\tau} - x^l_{v',t_\tau})^2 + (y^l_{v,t_\tau} - y^l_{v',t_\tau})^2} - D_v$ represents the distance between CAV $v$ and CAV $v'$ at lane $l$ at short-term time $t_\tau$, and $\Delta q^{\text{th}}$ is the distance threshold. We only consider the CAVs within this neighboring set. 

\subsection{Safety Metric}
Our CR model is based on the concept of TTC, which is a widely used metric in time-based safety measures. The TTC metric is employed to assess on-road safety, where each CAV continuously monitors its TTC with respect to neighboring vehicles \cite{1-choudhury2020experimental} and \cite{aznar2019time}. TTC is defined as the time required for two CAVs to reach a near-zero distance, which would result in a collision. Therefore, the TTC for CAV $v$ and the preceding CAV $(v-1)$ at lane $l$ at short-term time $t_\tau$ is computed as $\Xi^{l}_{v,t_\tau}=\Delta q^{l}_{v,t_\tau}/|\vartheta^l_{v,t_\tau}-\vartheta^{l}_{(v-1),t_\tau}|$,
where $\vartheta^l_{v,t_\tau} - \vartheta^{l}_{(v-1),t_\tau}$ is the relative velocity between CAV $v$ and the preceding CAV $(v-1)$ at lane $l$ at short-term time $t_\tau$. Overall, on-road safety can be measured by counting the number of instances between each pair of CAVs in which $\Xi^{l}_{v,t_\tau}$ falls below the threshold value of ${\Xi}^{\text{th}}$:
\begin{align}\label{eq-47}
	&CR=
	\begin{cases}
		1,&\text{if}~{\Xi}^{l}_{v,t_\tau}<\Xi^{\text{th}},\\
		0,&\text{otherwise}.
	\end{cases}
\end{align}

A collision occurs when ${\Xi}^{l}_{v,t_\tau}$ for each pair of CAVs exceeds the threshold value of $\Xi^{\text{th}} = \delta_{\text{React}} + \delta_{\text{Break}}$, where $\delta_{\text{React}}$ and $\delta_{\text{Break}}$ represent the time required to respond to the decision and to apply the brakes, respectively. In this paper, $\delta_{\text{React}}$ is assumed to be equal to 1s. For a CAV with speed $\vartheta^l_{v,t_\tau}$ and acceleration $z^l_{v,t_\tau}$, $\delta_{\text{Break}}$ is calculated as $\delta_{\text{Break}} = {\vartheta^l_{v,t_\tau}}/{z^l_{v,t_\tau}}$ \cite{ETSITR1033001,ETSITS1015393}. In order to determine the frequency at which the TTC exceeds the predetermined TTC threshold, we use the CR ratio, which serves as a measure of the proportion of instances that pose a collision risk (CR) to the total number of instances, both risky and non-risky, for each pair of CAVs. When considering the optimization problem objective for each CAV, the primary goal is to maximize ${\Xi}^{l}_{v,t_\tau}$ in order to reduce the CR ratio.

\subsection{Signal Model}
Assuming the use of Orthogonal Frequency-Division Multiple Access (OFDMA)\footnote{OFDMA enhances ISAC by reducing inter-user interference through subcarrier separation and improving sensing accuracy via efficient pilot designs \cite{11020763}.}, the total bandwidth $\Omega$ in the mmWave band is divided into $K$ subcarriers, indexed by $k$, with $\mathcal{K} = \{k_1, k_2, \dots, k_K\}$, each of bandwidth $w_k$. We further assume ideal, error-free backhaul links (i.e., low latency and high reliability) are employed to connect all RSUs to a CPU, facilitating coherent processing across the system \cite{yetis2021joint}. When considering the scenario with only one RSU, the primary challenge is intra-region interference, which arises due to energy leakage from the side beams [28]. Fortunately, intra-region interference can be mitigated by assigning orthogonal bandwidths to different vehicles. However, in a multi-RSU scenario, both intra-region and inter-region interference must be addressed. For example, consider CAV $v$ being served by RSU $r$, while CAV $v' \, (v' \neq v)$ is served by the neighboring RSU $r' \, (r' \neq r)$. At time slot $t_\tau$, if both vehicles move to nearby locations at the edges of different RSU service regions, beam collision may occur. Furthermore, if the allocated downlink bandwidths for CAVs $v$ and $v'$ overlap, inter-region interference will result. 

\subsubsection{Sensing Signal Model}
The transmitted signal vector by RSU $r$ to its serving CAVs $\forall v\in\mathcal{V}_r$ at short time slot $t_\tau$ on subcarrier $k$ is denoted by $\boldsymbol{\zeta}^k_{r,t_\tau} = [\zeta^k_{r1,t_\tau}, \zeta^k_{r2,t_\tau}, \dots, \zeta^k_{rv,t_\tau}, \dots, \zeta^k_{rV_r,t_\tau}]^T \in \mathbb{C}^{V_r \times 1}$, where $\zeta^k_{rv,t_\tau}$ represents the transmitted ISAC DL signal to vehicle $v$ at time slot $t_\tau$ on subcarrier $k$, whose power is assumed to be normalized, where $[.]^T$ denotes the transpose of a vector or matrix. Let $\textbf{F}^k_{r,t_\tau} = [\textbf{f}^k_{r1,t_\tau}, \textbf{f}^k_{r2,t_\tau}, \dots, \textbf{f}^k_{rv,t_\tau}, \dots, \textbf{f}^k_{rV_r,t_\tau}] \in \mathbb{C}^{M_t \times V_r}$ denote the transmit beamforming matrix at RSU $r$ at time slot $t_\tau$ on subcarrier $k$, where $\textbf{f}^k_{rv,t_\tau}$ is the beamforming vector for vehicle $v$ at time slot $t_\tau$ on subcarrier $k$ and $\mathbb{C}^{M_t \times V_r}$ denotes the set of all complex-valued matrices with $M_t$ rows and $V_r$ columns. Subsequently, the transmitted signal at RSU $r$ can be written as $\tilde{\boldsymbol{\zeta}}^k_{r,t_\tau} = \textbf{F}^k_{r,t_\tau} \boldsymbol{\zeta}^k_{r,t_\tau}=\sum_{v\in\mathcal{V}_r}\textbf{f}^k_{rv,t_\tau}\zeta^k_{rv,t_\tau} \in \mathbb{C}^{M_t \times 1}$.

In mmWave systems, the communication channel is typically represented by a Line-of-Sight (LoS) channel model \cite{10804675-1}. Due to the inherent sparsity of the mmWave channel, the Non-LoS (NLoS) component is significantly diminished, which facilitates more accurate channel modeling and subsequently enhances both sensing and communication capabilities. Through beam alignment, the transmitted signal associated with vehicle $v$ is reflected by the vehicle itself. Subsequently, RSU $r$ will receive $V_r$ reflected echoes at a given time slot $t_\tau$ on subcarrier $k$, which can be formulated as $\hat{\boldsymbol{\zeta}}^k_{r,t_\tau}= 
\alpha \sum_{v\in\mathcal{V}_r} \varrho^k_{rv,t_\tau}\textbf{b}^k(\theta_{rv,t_\tau})(\textbf{a}^k(\theta_{rv,t_\tau}))^H\tilde{\boldsymbol{\zeta}}^k_{r,t_\tau}(t_\tau-\tau_{rv,t_\tau})e^{j2\pi\tau_{rv,t_\tau}\nu^k_{rv,t_\tau}}+\textbf{n}^k_r,$
%\begin{align}\nonumber
%	&\hat{\boldsymbol{\zeta}}^k_{r,t_\tau}= 
%	\alpha \sum_{v\in\mathcal{V}_r} \varrho^k_{rv,t_\tau}\textbf{b}^k(\theta_{rv,t_\tau})(\textbf{a}^k(\theta_{rv,t_\tau}))^H\\&\times\tilde{\boldsymbol{\zeta}}^k_{r,t_\tau}(t_\tau-\tau_{rv,t_\tau})e^{j2\pi\tau_{rv,t_\tau}\nu^k_{rv,t_\tau}}+\textbf{n}^k_r,
%\end{align}
where $\alpha = \sqrt{M_t M_r}$ is the antenna array gain factor, $\textbf{n}_r\in \mathbb{C}^{M_r \times 1}$ denotes the complex Additive White Gaussian Noise (AWGN) vector, $\varrho^k_{rv,t_\tau} = \xi^k_{rv,t_\tau} (2 d_{rv,t_\tau})^{-1}$ is the reflection coefficient with a Radar Cross-Section (RCS) coefficient $\xi^k_{rv,t_\tau}$ and distance $d_{rv,t_\tau}$ \cite{10433790,10677488,10571114}, $\tau_{rv,t_\tau}$ and $\nu^k_{rv,t_\tau}$ denote the round-trip delay and the round-trip Doppler spread for vehicle $v$ at time slot $t_\tau$, respectively. Notably, RSUs positioned at different locations will capture distinct measurement parameters when observing the same CAV. Here, for RSU $r$, the azimuth angle,
elevation angle, and distance related to CAV $v$ at short time slot $t_\tau$ are denoted as $\theta_{rv,t_\tau}$, $\phi_{rv,t_\tau}$, and $d_{rv,t_\tau}$, respectively, in the Cartesian coordinate system. 

The terms $\textbf{b}(\theta_{rv,t_\tau}) \in \mathbb{C}^{M_r \times 1}$ and $\textbf{a}(\theta_{rv,t_\tau}) \in \mathbb{C}^{M_t \times 1}$ denote the receive and transmit steering vectors of RSU $r$, respectively, which are given by:
\begin{align}
	&\textbf{b}(\theta_{rv,t_\tau})=\sqrt{\frac{1}{M_r}}[1,\dots,e^{-j\pi(M_r-1)\cos\theta_{rv,t_\tau}}]^T,\\
	&\textbf{a}(\theta_{rv,t_\tau})=\sqrt{\frac{1}{M_t}}[1,\dots,e^{-j\pi(M_t-1)\cos\theta_{rv,t_\tau}}]^T,
\end{align}
respectively, where the assumption of half-wavelength antenna spacing is adopted.

\subsubsection{Radar Measurement Model}
In the considered V2I scenario enabled by a massive MIMO array, it is important to note that inter-beam interference between different vehicles in the uplink echoes can be effectively neglected when sufficiently narrow beams are employed. This allows the RSU to distinguish each vehicle based on the received echoes, as the steering vectors become asymptotically orthogonal \cite{10736521}. By utilizing spatial filtering techniques \cite{10484981}, the echo corresponding to vehicle $v$ at time slot $t_\tau$ can be isolated from the reflected signal echoes, which is mathematically expressed as $\hat{\zeta}^k_{rv,t_\tau}= \textbf{b}(\hat{\theta}_{rv,t_\tau})\hat{\boldsymbol{\zeta}}^k_{r,t_\tau}=
\alpha\varrho_{rv,t_\tau}\textbf{a}^{H}(\theta_{rv,t_\tau})\tilde{\boldsymbol{\zeta}}^k_{r,t_\tau}(t_\tau-\tau_{rv,t_\tau})e^{j2\pi\tau_{rv,t_\tau}\nu_{rv,t_\tau}}+{\hat{n}}^k_{rv,t_\tau},$
%\begin{align}\nonumber
%	&\hat{\zeta}^k_{rv,t_\tau}= \textbf{b}(\hat{\theta}_{rv,t_\tau})\hat{\boldsymbol{\zeta}}^k_{r,t_\tau}=
%	\alpha\varrho_{rv,t_\tau}\textbf{a}^{H}(\theta_{rv,t_\tau})\\&\times\tilde{\boldsymbol{\zeta}}^k_{r,t_\tau}(t_\tau-\tau_{rv,t_\tau})e^{j2\pi\tau_{rv,t_\tau}\nu_{rv,t_\tau}}+{\hat{n}}^k_{rv,t_\tau},
%\end{align}
where $\textbf{b}(\hat{\theta}_{rv,t_\tau})$ denotes the receive beamforming vector used for spatial filtering, with $\hat{\theta}_{rv,t_\tau}$ being an estimate of $\theta_{rv,t_\tau}$ obtained through existing Angle-of-Arrival (AoA) estimation techniques \cite{8770141} and $\textbf{a}^{H}$ denotes the Hermitian (conjugate transpose) of the complex vector $\textbf{a}$. It is important to note that the vehicle is modeled as a perfectly backscattering point target [40], thereby neglecting inter-vehicle reflections. Additionally, by allocating orthogonal bandwidths to vehicles within an RSU, the reflected echoes received by vehicles served by other RSUs can be guaranteed to remain orthogonal. This ensures the equivalence to mono-static sensing. For simplicity, we assume that $\hat{\theta}_{rv,t_\tau} = \theta_{rv,t_\tau}$, implying that $\textbf{b}^H(\hat{\theta}_{rv,t_\tau})\textbf{b}({\theta}_{rv,t_\tau}) = 1$ \cite{10745726}. The term $\hat{n}_{rv,t_\tau} = \textbf{b}^H(\hat{\theta}_{rv,t_\tau})\textbf{n}_r$ represents the complex AWGN with zero mean and variance $\sigma^2_k$. Additionally, the estimated delay $\tilde{\tau}_{rv,t_\tau}$ and Doppler shift $\tilde{\nu}_{rv,t_\tau}$ are acquired using the classic matched-filtering approach \cite{10061429}.
\begin{align}\label{eq-Vel-Delay}
	&\{\tilde{\tau}_{rv,t_\tau},\tilde{\nu}_{rv,t_\tau}\}=\nonumber
	\\&\arg\max_{\tau,\nu}\Big|\int_{0}^{\Delta T_d}\hat{\zeta}_{rv,t_\tau} \zeta^*_{rv,t_\tau}(t-\tau)e^{-j2\pi \nu_{rv,t_\tau} t}dt\Big|^2 ,
\end{align}
where $\Delta T_d$ represents the time duration of the received signals. Using the estimates $\tilde{\tau}_{rv,t_\tau}$ and $\tilde{\nu}_{rv,t_\tau}$ obtained from (\ref{eq-Vel-Delay}), and employing the interference cancellation method, we can effectively eliminate the multi-user interference. For simplicity, we assume ideal interference removal in this case. Consequently, we have $\bar{\zeta}_{rv,t_\tau}=\alpha\varrho_{rv,t_\tau}\textbf{a}^{H}(\theta_{rv,t_\tau})\textbf{f}^k_{rv,t_\tau}{{\zeta}}^k_{r,t_\tau}(t_\tau-\tau_{rv,t_\tau})e^{j2\pi\tau_{rv,t_\tau}\nu_{rv,t_\tau}}+{\hat{n}}^k_{rv,t_\tau}$. Thus, the measurement model for
angles $\theta_{rv,t_\tau}$ can be derived as
\begin{align}\label{eq-Vel-Delay-2}
	\tilde{\zeta}_{rv,t_\tau}&=\int_{0}^{\Delta T_d}\bar{\zeta}_{rv,t_\tau} \zeta^*_{rv,t_\tau}(t-\tilde{\tau})e^{-j2\pi \nu_{rv,t_\tau} t}dt\\&\nonumber=\alpha\varrho_{rv,t_\tau}G_m\textbf{a}^{H}(\theta_{rv,t_\tau})\textbf{f}^k_{rv,t_\tau}+{\tilde{n}}^k_{rv,t_\tau},
\end{align}
where $G_m$ is the matched-filtering gain and ${\tilde{n}}^k_{rv,t_\tau}$ represents the noise obeying the distribution $\mathcal{CN}\sim(0,\sigma^2_m)$ with variance $\sigma^2_m$. Based on the observation model, the distance $d_{rv,t_\tau}$ and radial velocity $\bar{\nu}_{rv,t_\tau}$ are given by:
\begin{align}
&d_{rv,t_\tau}=\frac{\tilde{\tau}_{rv,t_\tau}c}{2}=\frac{(\tilde{\tau}_{rv,t_\tau}+\epsilon_{rv,t_\tau})c}{2},\\
&\bar{\nu}_{rv,t_\tau}=\frac{\tilde{\nu}_{rv,t_\tau}c}{2}=\frac{(\tilde{\nu}_{rv,t_\tau}+\epsilon_{rv,t_\tau})c}{2},
\end{align}
where $f_c$ is the carrier frequency and $c$ is the velocity of signal propagation. The terms $\epsilon_{rv,t_\tau} \sim \mathcal{N}(0, \sigma^2_{rv,t_\tau})$ and $\tilde{\epsilon}_{rv,t_\tau} \sim \mathcal{N}(0, \tilde{\sigma}^2_{rv,t_\tau})$ represent the estimation errors of the distance and radial velocity for vehicle $v$, respectively. In particular, $\sigma^2_{rv,t_\tau}$ and $\tilde{\sigma}^2_{rv,t_\tau}$ typically depend on the Signal-to-Noise Ratios (SNRs) at the RSU and are given by:
\begin{align}
	&\sigma^2_{rv,t_\tau}=\frac{\rho^2\sigma^2_k}{G^2_m|\alpha\varrho_{rv,t_\tau}e^{-j2\pi \nu_{rv,t_\tau} t}|^2|\textbf{a}^{H}(\theta_{rv,t_\tau})\textbf{f}^k_{rv,t_\tau}|^2}\\
	&\tilde{\sigma}^2_{rv,t_\tau}=\frac{\tilde{\rho}^2\sigma^2_k}{G^2_m|\alpha\varrho_{rv,t_\tau}e^{-j2\pi \nu_{rv,t_\tau} t}|^2|\textbf{a}^{H}(\theta_{rv,t_\tau})\textbf{f}^k_{rv,t_\tau}|^2},
\end{align}
where $\rho$ and $\tilde{\rho}$ are constants determined by the specific system parameters. It is evident that the beamforming vector $\textbf{f}^k_{rv,t_\tau}$ significantly influences the noise variances $\sigma^2_{rv,t_\tau}$ and $\tilde{\sigma}^2_{rv,t_\tau}$. Consequently, a Deep Learning (DL)-based approach is employed to implicitly predict the state evolution and optimize the beamforming matrix $\textbf{f}^k_{rv,t_\tau}$, thereby enhancing estimation accuracy. 
\subsubsection{Communication Signal Model}
The communication signal received by vehicle $v$ is given by
\begin{align}\nonumber
&\check{\zeta}^k_{v,t_\tau}=\underbrace{\tilde{\alpha}\sum_{r\in \mathcal{R}_v}\textbf{a}^H({\theta}_{rv,t_\tau})\tilde{\varrho}^k_{rv,t_\tau}\sum_{v\in \mathcal{V}_r}\textbf{f}^k_{rv,t_\tau}\zeta^k_{rv,t_\tau}e^{j2\pi\nu_{rv,t_\tau}t}}_{\text{Desired signal}}
\end{align}

\begin{align}\nonumber&
=\underbrace{\tilde{\alpha}\sum_{v\prime\in \mathcal{V}_r\setminus v}\sum_{r\in \mathcal{R}_{v\prime}}\textbf{a}^H({\theta}_{rv,t_\tau})\tilde{\varrho}^k_{rv,t_\tau}\sum_{v\in \mathcal{V}_r}\textbf{f}^k_{rv,t_\tau}\zeta^k_{rv,t_\tau}e^{j2\pi\nu_{rv,t_\tau}t}}_{\text{Interference signal}}
\\&+w_{v,t_\tau},
\end{align}
where $\tilde{\alpha}^2=M_t$ represents the antenna gain and $\tilde{\varrho}_{rv,t_\tau}$ represents the communication channel coefficient. Additionally, $w_{rv,t_\tau}$ denotes the noise term at CAV $v$, which follows a complex normal distribution $\mathcal{CN}(0,\sigma_c^2)$ with an associated noise variance $\sigma_c^2$. Specifically, the channel coefficient $\tilde{\varrho}_{rv,t_\tau}$ can be expressed as $\tilde{\varrho}_{rv,t_\tau}=(\check{\varrho}_{rv,t_\tau}/d_{rv,t_\tau})e^{j2\pi (f_c/c) d_{rv,t_\tau}}$ [28], where, $\check{\varrho}_{rv,t_\tau}$ represents the channel gain constant at the reference distance $d_0=1$. Given that $\check{\varrho}_{rv,t_\tau}$, $f_c$, and $c$ are known at the RSU, the channel coefficient $\tilde{\varrho}_{rv,t_\tau}$ can be determined based on $d_{rv,t_\tau}$.

Assuming that the transmitted communication signal has unit power, the received Signal-to-Interference-plus-Noise Ratio (SINR) for vehicle $v$ at time slot is given by \cite{10571114}
\begin{align}
	&\psi^k_{rv,t_\tau}(\textbf{h}^k_{rv,t_\tau},\textbf{F}^k_{r,t_\tau})= \\\nonumber&\frac{\sum_{r\in \mathcal{R}_v}|(\textbf{h}^k_{rv,t_\tau})^H\textbf{f}^k_{rv,t_\tau}|^2}{\sum_{v\prime\in \mathcal{V}_r\setminus v}|\sum_{r\in \mathcal{R}_{v\prime}}(\textbf{h}^k_{rv\prime,t_\tau})^H\textbf{f}_{rv\prime,t_\tau}|^2+\sigma^2_c},
\end{align}
where $\textbf{h}^k_{rv,t_\tau}$ denotes the equivalent channel vector for vehicle $v$ at time slot $t_\tau$. Consequently, the achievable rate $R_{rv,t_\tau}$ for vehicle $v$ at time slot $t_\tau$ can be expressed as $R^k_{rv,t_\tau} (\textbf{h}_{rv,t_\tau}, \textbf{F}_{r,t_\tau})) = \log_2(1 + \psi^k_{rv,t_\tau}(\textbf{h}_{rv,t_\tau}, \textbf{F}_{r,t_\tau}))$. In this context, multi-user interference from other vehicles is considered during the sensing process, specifically $\sum_{v' \neq v} \textbf{f}_{rv',t_\tau} \zeta^k_{rv,t_\tau}$. Similar to the sensing task, the achievable sum-rate $\sum_{v} R_{rv,t_\tau}$ can be enhanced by adapting the beamforming matrix $\textbf{F}_{r,t_\tau}$. The channel capacity of V2I links must satisfy the following constraints, respectively:
\begin{align}\label{eq-16}
	&\sum_{k\in\mathcal{K}}\rho^k_{rv,t_\tau}R^k_{rv,t_\tau}(\textbf{h}_{rv,t_\tau}, \textbf{F}_{r,t_\tau})\geq \bar{C}_{v},\forall v,t_\tau, 
\end{align}
where $\bar{C}_{v}$ is the minimum required rate for receiver CAV $v$ to receive the requested exogenous information. 

Based on the safety requirement, each RSU $r$ utilizes a set of the binary variables $\beta_{rvv',t_\tau}\in\{0,1\}$ to decide to sense HDV at short-term time $t_\tau$ and send their sensing information to CAV $v'$ at coverage area or not. This will be established by setting  $\beta_{rvv',t_\tau}=1$ and $\beta_{rvv',t_\tau}=0$, respectively. 

Using the motion model, the relationship between the estimated parameters $(d_{rv,t_\tau},\theta_{rv,t_\tau})$ and the vehicle’s position $({x}^l_{v,t_\tau},{y}^l_{v,t_\tau})$ can be expressed as ${d}_{rv,t_\tau}=\sqrt{({x}^l_{v,t_\tau})^2 + ({y}^l_{v,t_\tau})^2}$, and ${\theta}_{rv,t_\tau}=\arctan\left({y}^l_{v,t_\tau}/{x}^l_{v,t_\tau}\right).$ As shown in \([32]\), the kinematic model characterizes the changes in distance and angle for CAV $v$ as $d^2_{rv,t_\tau}=d^2_{rv,t_\tau-1}+\Delta d^2 - 2d^2_{rv,t_\tau-1} \Delta d \cos\theta_{rv,t_\tau-1}, \Delta d \sin\theta_{rv,t_\tau-1}=d_{n} \sin\Delta\theta$, where $\Delta d$ and $\Delta \theta$ represent the distance and angle variations over one time slot. The beam tracking process focuses on monitoring these changes in distances and angles for $V$ vehicles using the received signals defined in (\ref{eq-Vel-Delay-2}).  

\section{Value of Status and Augmented-state Sequential Stochastic Decision Process (SSDP)}\label{Value-Info}
%In an ideal situation, RSUs would be capable of instantly detecting the position and speed of all HDVs and transmitting this data to CAVs to improve road safety. However, real-world networks face significant challenges, such as limited bandwidth and delays caused by restricted onboard processing power. Our research focuses on prioritizing critical safety-related information to reduce the overall communication load, addressing gaps not fully considered in earlier studies. By optimizing the exchange of essential data, we aim to decrease transmission delays and improve reliability. Techniques like data compression help minimize the amount of transmitted data, allowing networks to accommodate more vehicles. However, cellular systems, such as 5G, operate with a finite number of Resource Blocks (RBs) per Transmission Time Interval (TTI). For example, a 20 MHz 5G channel supports 100 RBs. When the number of vehicles exceeds this capacity, the network cannot serve all vehicles simultaneously. While data compression alleviates some strain, high vehicle density still limits the network’s ability to transmit data efficiently, potentially leading to the transmission of non-critical data at the expense of crucial safety information. 
In an ideal scenario, RSUs could instantly detect HDV positions and speeds and share them with CAVs to enhance safety. In practice, network constraints—such as limited bandwidth, processing delays, and finite 5G resource blocks—restrict this capability. Our research addresses these limitations by prioritizing critical safety data, thereby reducing communication load, minimizing delays, and improving reliability. Although methods like data compression ease congestion, high vehicle density still risks transmitting non-essential data at the expense of vital safety information.

%We consider that each CAV independently selects its movement and radio channel actions, denoted as $\textbf{u}_{v}$ and $\boldsymbol{\alpha}_{v}$, based on data gathered from its onboard sensors and V2X communication ${\boldsymbol{\beta}}_{v}$. While incorporating information from other CAVs—referred to as exogenous information—has the potential to enhance both transportation efficiency and radio network performance, it can also result in significant drawbacks, such as increased signaling overhead, heavier network traffic, higher costs, and computational challenges due to the curse of dimensionality. Thus, there is a necessary balance between the benefits of using exogenous information and the associated performance trade-offs. Selecting an appropriate subset of this information is essential to optimize system performance. Specifically, the chosen subset should represent the most valuable exogenous information available. To achieve this, each CAV identifies the CAVs whose shared data would most effectively assist it in making improved decisions.
Each CAV determines its movement and radio channel actions, denoted by $\textbf{u}{v}$ and $\boldsymbol{\alpha}{v}$, using onboard sensing and V2X communication $\boldsymbol{\beta}_{v}$. Incorporating exogenous information from other CAVs can improve transportation efficiency and network performance but also introduces challenges such as signaling overhead, network congestion, higher costs, and computational complexity from the curse of dimensionality. Hence, a balance is required between the benefits and trade-offs of using such information. Optimizing system performance relies on selecting the most valuable subset of exogenous information, i.e., the data from CAVs whose contributions are most beneficial for decision-making.

Each CAV $v$ has local information about movement at large-term and short-term times, $s^{\text{L},l}_{v,\tau}=\{u^{l}_{v,\tau},\alpha^{l}_{v,\tau}\}$ and $s^{\text{S},l}_{v,t_{\tau}}=\{e^{l}_{v,t_{\tau}},\tilde{e}^{l}_{v,t_{\tau}},{\vartheta}^l_{v,t_\tau},{z}^l_{v,t_\tau},{\theta}^l_{v,t_\tau},\textbf{q}^{l}_{v,t_\tau}\}$, respectively. However, in order to improve the performance of both transportation and communication networks, each CAV needs to obtain the information of other CAVs (i.e., exogenous information), large-term time information $\bar{s}^{\text{L},l}_{v,\tau}=\{s^{\text{L},l}_{v',\tau}\}_{\substack{v'\in\mathcal{V}, v'\neq v}}=\{u^{l}_{v',\tau},\alpha^{l}_{v',\tau}\}_{\substack{v'\in\mathcal{V}, v'\neq v}}$ and short-term time information $\bar{s}^{\text{S},l}_{v,\tau}=\{s^{\text{S},l}_{v',t_\tau}\}_{\substack{v'\in\mathcal{V}, v'\neq v}}=\{e^{l}_{v',t_{\tau}},\tilde{e}^{l}_{v',t_{\tau}},{\vartheta}^l_{v',t_\tau},{z}^l_{v',t_\tau},{\theta}^l_{v',t_\tau},\textbf{q}^{l}_{v',t_\tau}\}_{v',v"\in\mathcal{V}, v'\neq v}$. In the next subsection, after introducing SSDP and augmented-state SSDP formulations, we adopt the KL divergence measure to select a subset of all available states that can help each CAV to improve its controlling policy in selecting movement and radio parameters.

\subsection{SSDP and Augmented-state SSDP Formulation}
\begin{definition}
A standard SSDP over a finite time horizon $t \in \{t_\tau, \tau\}$ is characterized for each agent $k$ by the elements $\{s_{k,t}, a_{k,t}, \bar{s}_{k,t}, f^s, f^{\bar{s}}, r_{k,t}\}$. Here, $s_{k,t} \in \mathcal{S}$ represents the state of agent $k$ at time $t$, where $\mathcal{S}$ is the state space. Similarly, $a_{k,t} \in \mathcal{A}$ denotes the action taken at time $t$, with $\mathcal{A}$ being the action space. The exogenous information at time $t$, represented as $\bar{s}_{k,t} \in \bar{\mathcal{S}}$, lies within the outcome space $\bar{\mathcal{S}}$ and becomes available after performing action $a_{k,t}$. 

The functions $f^s$ and $f^{\bar{s}}$ define the state transition and the evolution of exogenous information, respectively. The next state is determined by the state transition function as $s_{k,t+1} = f^s(s_{k,t}, a_{k,t}, \bar{s}_{k,t})$, while the next exogenous information is computed as $\bar{s}_{k,t+1} = f^{\bar{s}}(\{s_{k,t'}\}_{t'=0}^{t+1}, \{a_{t'}\}_{t'=0}^{t+1}, \{\bar{s}_{k,t'}\}_{t'=0}^{t}, \xi_{k,t})$. Here, $\xi_{k,t}$ accounts for all factors influencing $\bar{s}_{k,t+1}$ aside from the states, actions, and prior exogenous information up to time $t+1$. 

The reward function at time $t$ is denoted as $r_{k,t}(s_{k,t}, a_{k,t}, \bar{s}_{k,t})$, and for each time step, a policy $\pi_k$ is defined as a set of functions $\mu_{k,t}$ that maps the current state to an action, expressed as $a_{k,t} = \mu_k(s_{k,t})$. Under a given policy $\pi_k$, the expected cumulative reward over the finite time horizon can be calculated. The objective is to identify the optimal policy $\pi^*_k$ that maximizes this total expected reward.
\end{definition}

In an SSDP, during each time step $t$, the action $a_{k,t}$ is determined solely by the current state $s_{k,t}$, without any prior information about the exogenous variable $\bar{s}_{k,t}$.
\begin{definition}
	When the exogenous information $\bar{s}_{k,t}$ is known prior to taking action $a_{k,t}$, an augmented-state SSDP over a finite time horizon ${t} \in \mathcal{T}$ can be represented by $\{\tilde{s}_{k,t}, a_{k,t}, f^{\tilde{s}}, f^{\bar{s}}, r_{k,t}\}$. In this formulation, the augmented state $\tilde{s}_{k,t} = (s_{k,t}, \bar{s}_{k,t}) \in \mathcal{S}$ is created by incorporating the exogenous information $\bar{s}_{k,t}$ into the original state space of the standard SSDP. 
	
	The action $a_{k,t} \in \mathcal{A}$ and the reward function $r_{k,t}(s_{k,t}, a_{k,t}, \bar{s}_{k,t}) = r_{k,t}(\tilde{s}_{k,t}, a_{k,t})$ remain unchanged from those in the original SSDP. The exogenous information $\bar{s}_{k,t}$ at time $t$ is determined as $\bar{s}_{k,t} = \{\{s_{k,t'}\}_{t'=0}^{t-1}, \{a_{k,t'}\}_{t'=0}^{t-1}, \{\bar{s}_{k,t'}\}_{t'=0}^{t-1}, \xi_{k,t}\}$, where $\xi_{k,t}$ encompasses additional parameters influencing $\bar{s}_{k,t}$. 
	
	In this case, the state transition function $f^s$ is adapted to reflect the augmented state structure.
	\begin{align}
		&\tilde{s}_{k,{t}+1}=f^{\tilde{s}}\left({s}_{k,t},a_{k,t},\bar{s}_{k,t}\right)
		=\begin{pmatrix} 
			s_{k,{t}+1} \\ 
			\bar{s}_{k,{t}+1}  
		\end{pmatrix}\\&\nonumber=\begin{pmatrix} 
			f^s\left(s_{k,t},a_{k,t},\bar{s}_{k,t}\right) \\ 
			f^{\bar{s}}\left(\{s_{k,{t'}}\}_{{t'}=0}^{{t}+1},\{a_{k,t'}\}_{{t'}=0}^{{t}+1},\{\bar{s}_{k,t'}\}_{{t'}=0}^{{t}}, \xi_{k,t}\right) 
		\end{pmatrix}.
	\end{align}
	
Likewise, at each time step $t$, a policy $\tilde{\pi}_k$ specifies the action for a given state, such that the action $a_{k,t}$ is chosen according to $\tilde{\mu}_k(\tilde{s}_{k,t})$.
\end{definition}

\begin{remark}
	Definition 1 outlines the SSDP framework, which includes the most general form of the exogenous information transition function $f^{\bar{s}}$. However, by placing a restriction on $f^{\bar{s}}$, such that the exogenous information transition is defined as $\bar{s}_{k,t+1} = f^{\bar{s}}(s_{k,t+1}, a_{k,t+1}, \xi_{k,t})$, where $\xi_{k,t}$ is an independent random variable following a predefined distribution, the SSDP effectively reduces to an MDP.
\end{remark}

\subsection{The Value of Exogenous Information Analysis}
In this subsection, we focus on determining $\tilde{\pi}^*_k(\tilde{s}_{k,t})$, which leverages $\bar{s}_{k,t}$ to achieve better performance compared to $\pi^*_k(s_{k,t})$. It is important to highlight that $\tilde{\pi}^*_k(\tilde{s}_{k,t})$ and $\pi^*_k(s_{k,t})$ will yield identical results if $\bar{s}_{k,t}$ has no influence on the reward function or the state transition dynamics of $s_{k,t}$. Consequently, the performance of $\tilde{\pi}^*_k(\tilde{s}_{k,t})$ is guaranteed to be at least as effective as that of $\pi^*_k(s_{k,t})$. The significance of $\bar{s}_{k,t}$ and its impact on $\tilde{\pi}^*_k(\tilde{s}_{k,t})$ are determined by how $\bar{s}_{k,t}$ affects both the reward function and the state transition. Therefore, the optimal policy $\tilde{\pi}^*_k(\tilde{s}_{k,t})$ for the augmented-state SSDP either outperforms or matches the original SSDP policy $\pi^*_k(s_{k,t})$, provided that the exogenous information evolves according to $\bar{s}_{k,t+1}=f^{\bar{s}}(s_{k,t},\bar{s}_{k,t},\xi_{k,t})$.

\begin{remark}
The optimal policy of the augmented-state SSDP, represented as $\tilde{\pi}^{*}_k(\tilde{s}_{k,t})$, demonstrates performance that is either equivalent to or better than the optimal policy of the original SSDP, $\pi^*_k(s_{k,t})$, provided that the exogenous information adheres to the condition $\bar{s}_{k,t+1} = f^{\bar{s}}(s_{k,t+1}, \xi_{k,t})$. In this scenario, the exogenous information $\bar{s}_{k,t+1}$ is allowed to depend on the state $s_{k,t+1}$, but it must not depend on any other variables aside from $\xi_{k,t}$, which is a random variable with an independent distribution.
\end{remark}

In the following, we model the effect of $\bar{s}_{k,t}$ on state transitions as the question: "How much more accurately can the state $s_{k,{t}+1}$ be predicted when $\tilde{s}_{k,t}$ incorporates $\bar{s}_{k,t}$?" To explore this, we transform the transition functions for the system state and exogenous information, namely $f^{s}$, $f^{\tilde{s}}$, and $f^{\bar{s}}$, into their corresponding transition probabilities: $T^s = \Pr\{s_{k,{t}+1}|s_{k,t},a_{k,t}\}$, $T^{\tilde{s}} = \Pr\{s_{k,{t}+1},\bar{s}_{k,{t}+1}|s_{k,t},a_{k,t},\bar{s}_{k,t}\}$, and $T^{\bar{s}} = \Pr\{\bar{s}_{k,{t}+1}|\bar{s}_{k,t},s_{k,t},a_{k,t}\}$. The conditional KL divergence between $T^s \otimes T^{\bar{s}}$ and $T^{\tilde{s}}$ can then be computed as follows:
\begin{align}\nonumber
	&\mathcal{DL}(T^{\tilde{s}}\parallel T^s \otimes T^{\bar{s}})=\int_{\tilde{s}_{k,{t}+1},\tilde{s}_{k,t},a_{k,t}}\Pr\{\tilde{s}_{k,{t}+1},\tilde{s}_{k,t},a_{k,t}\}\\&\times\log_2\Big(\frac{\Pr\{\tilde{s}_{k,{t}+1}|\tilde{s}_{k,t},a_{k,t}\}}{\Pr\{s_{k,{t}+1}|s_{k,t},a_{k,t}\}\Pr\{\bar{s}_{k,{t}+1}|s_{k,t},a_{k,t},\bar{s}_{k,t}\}}\Big),
\end{align}
where $\otimes$ represents the Kronecker product. When the transition of $s_{k,t}$ is independent of $\bar{s}_{k,t}$, the KL divergence equals zero, meaning that $\bar{s}_{k,t}$ does not contribute to the transition dynamics of $s_{k,t}$, and thus is unnecessary. Conversely, a higher KL divergence indicates that incorporating $\bar{s}_{k,t}$ can enhance the accuracy of predicting the future state. By evaluating the KL divergence, each CAV can determine which specific subset of exogenous information would be most effective for improving control policies. In the following sections, we explore how exogenous information contributes to both transportation and radio network control policies. Given that CAV mechanical systems tend to operate slower than communication systems, a two-time-scale DDPG algorithm is employed to reduce signal processing complexity, minimize system load, and cut down on signaling overhead, all while preserving performance. 

Our method utilizes a two-time-scale framework to reduce the complexity of signal processing and the overhead of signaling. In this framework, the value of exogenous information is categorized into two components: long-term (denoted as $\text{L}$) and short-term (denoted as $\text{S}$). The long-term component includes information related to the CAV's control system and the radio network, which is transmitted over extended periods, denoted by times $\tau$ and $t_\tau$, respectively.
\subsubsection{Value of Exogenous Information for Large-term Time}
Although the transmission of all large-term time exogenous information can improve the optimal control policy of CAV $v$, the transmission of this information $\bar{s}^{\text{L},l}_{v,\tau}=\{s^{\text{L},l'}_{v',\tau}\}_{\substack{\l'\in\mathcal{L},v'\in\mathcal{V}, v'\neq v}}=\{u^{l'}_{v',\tau},\alpha^{l'}_{v',\tau}\}_{\substack{\l',l"\in\mathcal{L},v',v"\in\mathcal{V}, v'\neq v}}$ can lead to the performance degradation due to the communication and computation overheads, and the curse-of-dimensionality. Therefore, a trade-off between improving the performance of the optimal policy and reducing the state space dimension should be considered by transmitting only the high value components that help  better predict the future state at CAV $v$. In other words, only a subset of exogenous information is used for predicting future states in the augmented-state to get improved DRL-based control policy.
To this end, the value of each exogenous information is quantified. For each CAV $v$, the augmented-state $\tilde{s}^{\text{L},l}_{v,\tau}$ can include its driving status $s^{\text{L},l}_{v,\tau}=\{u^{l}_{v,\tau},\alpha^{l}_{v,\tau}\}$ and exogenous information $\bar{s}^{\text{L},l}_{v,\tau}=s^{\text{L}}_{\tau}\setminus s^{\text{L},l}_{v,\tau}$ transmitted by V2X from other CAVs, where $s^{\text{L}}_{\tau}=\{s^{\text{L},l}_{v,\tau},\{s^{\text{L},l'}_{v',\tau}\}_{l'\in\mathcal{L},v'\in\mathcal{V},v'\neq v}\}$. If the state $\hat{s}^{\text{L},l}_{v,\tau}$ is a subset of all exogenous information $\bar{s}^{\text{L},l}_{v,\tau}$ that can be used by CAV $v$ to predict the future actions of its preceding CAV ${(v-1)}$, the
value of this additional information $\hat{s}^{\text{L},l}_{v,\tau}$ for CAV $v$ can be analyzed by deriving the KL divergence for including the additional information as (\ref{KL-eq-1})
\begin{table*}
	\centering
	\begin{minipage}{\textwidth}
		\small
		\begin{align}\label{KL-eq-1}\nonumber
			&\mathcal{DL}^{\text{L},l}_{vv'}\Big(T^{\tilde{s}^{\text{L},l}_{v,\tau}}\parallel T^{s^{\text{L},l}_{v,\tau}} \otimes T^{s^{\text{L},l'}_{v',\tau}}\Big)=	\int_{\tilde{s}^{\text{L},l}_{v,(\tau+1)},\tilde{s}^{\text{L},l}_{v,\tau},a^{\text{L},l}_{v,\tau}}\Pr\{\tilde{s}^{\text{L},l}_{v,(\tau+1)},\tilde{s}^{\text{L},l}_{v,\tau},a^{\text{L},l}_{v,\tau}\}\log_2\Bigg(\frac{\Pr\{\tilde{s}^{\text{L},l}_{v,(\tau+1)}|\tilde{s}^{\text{L},l}_{v,\tau},a^{\text{L},l}_{v,\tau}\}}{\Pr\{s^{\text{L},l}_{v,(\tau+1)}|s^{\text{L},l}_{v,\tau},a^{\text{L},l}_{v,\tau}\}\Pr\{s^{\text{L},l'}_{v',(\tau+1)}|s^{\text{L},l}_{v,\tau},s^{\text{L},l'}_{v',\tau}\}}\Bigg)
			\\&=\int_{\substack{a^{\text{L},l}_{(v-1),(\tau+1)}}}\Pr\{a^{\text{L},l}_{(v-1),(\tau+1)},s^{\text{L},l'}_{v',\tau}\}\log_2\Bigg(\frac{\Pr\{a^{\text{L},l}_{(v-1),(\tau+1)}|s^{\text{L},l'}_{v',\tau}\}}{\Pr\{a^{\text{L},l}_{(v-1),(\tau+1)}|s^{\text{L},l}_{v,\tau}\}}\Bigg),
		\end{align}
		%		\medskip
		%		\hrule
	\end{minipage}
\end{table*}
where $a^{\text{L},l}_{(v-1),(\tau+1)}=\{u^{l}_{(v-1),(\tau+1)},\alpha^{l}_{(v-1),(\tau+1)}\}$. As can be seen from (\ref{KL-eq-1}), this KL divergence depends on the ratio of  $\Pr\{u^{l}_{(v-1),(\tau+1)},\alpha^{l}_{(v-1),(\tau+1)}|\hat{s}^{\text{L},l}_{v,\tau}\}$ and $\Pr\{u^{l}_{(v-1),(\tau+1)},\alpha^{l}_{(v-1),(\tau+1)}|s^{\text{L},l}_v(t_\tau)\}$ and shows how much better CAV $v$ can predict the control parameters $u^{l}_{(v-1),(\tau+1)}$ and $\alpha^{l}_{(v-1),(\tau+1)}$ of CAV ${(v-1)}$ in large-term time $\tau+1$, given the additional information from CAV ${v'}$. Therefore, we form $\hat{s}^{\text{L},l}_{v,\tau}$ as $
	\hat{s}^{\text{L},l}_{v,\tau}=\left\lbrace s^{\text{L},l'}_{v',\tau}|\mathcal{DL}^{\text{L},l}_{vv'}\geq\mathcal{DL}^{\text{th}}_v\right\rbrace$ , where $\mathcal{DL}^{\text{th}}_v$ denotes the threshold for determining the high value information. We deploy Monte Carlo (MC) method to estimate the KL divergence, i.e., the value of (\ref{KL-eq-1}) numerically,
%and (\ref{KL-eq}) 
as the expected value of $\log_2\Big(\frac{\Pr\{a^{\text{L},l}_{(v-1),(\tau+1)}|s^{\text{L},l'}_{v',\tau}\}}{\Pr\{a^{\text{L},l}_{(v-1),(\tau+1)}|s^{\text{L},l}_{v,\tau}\}}\Big)$.
By generating $F$ independent and identically distributed (i.i.d) samples from  $\Pr\{a^{\text{L},l}_{(v-1),(\tau+1)},s^{\text{L},l'}_{v',\tau}\}$, the approach involves computing $\mathcal{DL}^{\text{L},l}_{vv'}=\frac{1}{F}\sum_{i=1}^{F}\log_2\Big(\frac{\Pr\{a^{\text{L},l}_{(v-1),(\tau+1)}|s^{\text{L},l'}_{v',\tau}\}}{\Pr\{a^{\text{L},l}_{(v-1),(\tau+1)}|s^{\text{L},l}_{v,\tau}\}}\Big)$.

As the number of samples, $F$, increases, the MC estimation error decreases. With such an approximation, the estimation error distribution is normal with zero mean and variance ${\sigma}^{\text{L},l}_{vv'}$, where ${\sigma}^{\text{L},l}_{vv'}=\frac{1}{F}\mathbb{E}\{[\log_2\Big(\frac{\Pr\{a^{\text{L},l}_{(v-1),(\tau+1)}|s^{\text{L},l'}_{v',\tau}\}}{\Pr\{a^{\text{L},l}_{(v-1),(\tau+1)}|s^{\text{L},l}_{v,\tau}\}}\Big)]^2\}$, where $\mathbb{E}\{.\}$ the expected value (mathematical expectation).

Based on the KL divergence value, CAV $v$ decides to request the state information of CAV $v'$ in order to better predict the preceding CAV $(v-1)$ actions.

\subsubsection{Value of Exogenous Information for Short-term Time}
A similar way can be considered for calculating the value of exogenous information of the radio channel. If the state $s^{\text{S},l}_{v,t_\tau}$ is a subset of the exogenous information of channel $\bar{s}^{\text{S},l}_{v,t_\tau}$ that can be used by CAV $v$ to predict the future state of CAV ${(v-1)}$, the
value of this additional information $s^{\text{S},l}_{v,t_\tau}$ for CAV $v$ can be analyzed by deriving the KL divergence for including the additional information as (\ref{KL-eq})
\begin{table*}
	\centering
	\begin{minipage}{\textwidth}
		\small
		\begin{align}\label{KL-eq}
			&\mathcal{DL}^{\text{S},l}_{vv'}\Big(T^{\tilde{s}^{\text{S},l}_{v,t_\tau}}\parallel T^{s^{\text{S},l}_{v,t_\tau}} \otimes T^{s^{\text{S},l'}_{v',t_\tau}}\Big)=	\int_{\tilde{s}^{\text{S},l}_{v,(t_\tau+1)},\tilde{s}^{\text{S},l}_{v,t_\tau},a^{\text{S},l}_{v,t_\tau}}\Pr\{\tilde{s}^{\text{S},l}_{v,(t_\tau+1)},\tilde{s}^{\text{S},l}_{v,t_\tau},a^{\text{S},l}_{v,t_\tau}\}\times\\\nonumber&\log_2\Bigg(\frac{\Pr\{\tilde{s}^{\text{S},l}_{v,(t_\tau+1)}|\tilde{s}^{\text{S},l}_{v,t_\tau},a^{\text{S},l}_{v,t_\tau}\}}{\Pr\{s^{\text{S},l}_{v,(t_\tau+1)}|s^{\text{S},l}_{v,t_\tau},a^{\text{S},l}_{v,t_\tau}\}\Pr\{s^{\text{S},l'}_{v',(t_\tau+1)}|s^{l}_{v,t_\tau},s^{\text{S},l'}_{v',t_\tau}\}}\Bigg)
			=\int_{\substack{a^{\text{S},l}_{(v-1),(t_\tau+1)}}}\Pr\{a^{\text{S},l}_{(v-1),(t_\tau+1)},s^{\text{S},l'}_{v',t_\tau}\}\log_2\Bigg(\frac{\Pr\{a^{\text{S},l}_{(v-1),(t_\tau+1)}|s^{\text{S},l'}_{v',t_\tau}\}}{\Pr\{a^{\text{S},l}_{(v-1),(t_\tau+1)}|s^{\text{S},l}_{v,t_\tau}\}}\Bigg).
		\end{align}
	\end{minipage}
\end{table*}
where $a^{\text{S},l}_{v,t_\tau}=\{\beta_{rv'v,t_\tau}\}$. Therefore, we form $\hat{s}^{\text{S},l}_{v,t_\tau}$ as 
\begin{align}
	\hat{s}^{\text{S},l}_{v,t_\tau}=\left\lbrace s^{\text{S},l'}_{v',t_\tau}|\mathcal{DL}^{\text{S},l}_{vv'}\geq\mathcal{DL}^{\text{th}}_v\right\rbrace .
\end{align}
We take similar steps to deal with the value of (\ref{KL-eq}) based on MC.
Note that each RSU has only the short-time term state and action that can be determined as $\textbf{s}^{\text{S}}_{r,t_\tau}=\{{\vartheta}^l_{v,t_\tau},{z}^l_{v,t_\tau},{\theta}^l_{v,t_\tau},\textbf{q}^{l}_{v,t_\tau},\textbf{h}^k_{rv,t_\tau}\}$ and $\textbf{a}^{\text{S}}_{r,t_\tau}=\{\textbf{f}^k_{rv,t_\tau}\}$.
\section{Problem Formulation for Radio Resource Allocation and Vehicle Control}\label{Problem Formulation}
We propose a predictive DL-based beamforming framework that optimizes sensing while ensuring DL communication quality, reducing signaling overhead, and improving robustness. Performance is evaluated using Cramer-Rao Bounds (CRBs) and an associated optimization problem.

\subsection{Cramer-Rao Bound for Sensing Performance}
The CRB is commonly used to assess the accuracy of parameter estimation. Initially, we compute the Fisher Information Matrix (FIM) based on the CRB theorem, as outlined in [34]
\begin{align}
	\textbf{J}(o^k_{rv,t})=\mathbb{E}\left\lbrace \left[\frac{\partial \ln p(\tilde{o}^k_{rv,t},o^k_{rv,t})}{\partial o^k_{rv,t}} \right] \left[\frac{\partial \ln p(\tilde{o}^k_{rv,t},o^k_{rv,t})}{\partial o^k_{rv,t}} \right]^T \right\rbrace,
\end{align}
where $\ln p(\tilde{o}^k_{rv,t},o^k_{rv,t})$ is the likelihood function of $\tilde{o}^k_{rv,t}=[\tilde{\zeta}_{rv,t_\tau},  \tilde{\tau}_{rv,t_\tau},\tilde{\nu}_{rv,t_\tau}]^T$ conditioned on the motion parameter
vector $o^k_{rv,t}=[\theta_{rv,t_\tau},d_{rv,t_\tau},\vartheta_{v,t_\tau}]^T$.
In this scenario, the relationship is expressed as $\tilde{o}^k_{rv,t}=\phi(o^k_{rv,t})+\bar{n}^k_{rv,t}$, where $\phi(\cdot)$ represents the nonlinear function of the observation vector, and $\bar{n}^k_{rv,t}=[\tilde{n}_{k,n}, \epsilon_{k,n}, \varepsilon_{k,n}]$ includes the noise components. Consequently, $\tilde{o}^k_{rv,t}$ follows a complex Gaussian distribution, $\tilde{o}^k_{rv,t}\sim \mathcal{CN}(\phi(o^k_{rv,t}), \Sigma)$, with a covariance matrix $\Sigma = \text{diag}(\sigma_n^2 1_{M_r}, \sigma_\tau^2, \sigma_\nu^2)$. Based on this, the conditional Probability Density Function (PDF) can be derived as follows
\begin{align}
	&p(\tilde{o}^k_{rv,t},o^k_{rv,t})
	\\&\nonumber=\frac{1}{\pi^{M_r+2}\text{det}(\Sigma)}e^{-(\tilde{o}^k_{rv,t}-\phi(o^k_{rv,t}))^H\Sigma^{-1}(\tilde{o}^k_{rv,t}-\phi(o^k_{rv,t}))}.
\end{align}

Once the FIM is obtained, the CRBs for the parameters $\theta_{rv,t}$ and $d_{rv,t}$, corresponding to vehicle $v$ at time slot $t$, can be expressed as follows:
\begin{align}
	&\Phi(d^k_{rv,t},f^k_{rv,t})=\left[\frac{1}{\sigma^2_\tau} \left(\frac{2}{c} \right)^2 \right]^{-1},
	\\&\Phi(\theta^k_{rv,t},f^k_{rv,t})=\left[\frac{1}{\sigma^2_\tau} \left(\frac{\partial \tilde{\zeta}_{rv,t_\tau}}{\partial \theta^k_{rv,t}} \right)^2 \left(\frac{\partial \tilde{\zeta}_{rv,t_\tau}}{\partial \theta^k_{rv,t}} \right)^2 \right]^{-1}.
\end{align}

The Mean Squared Error (MSE) matrix of $o^k_{rv,t}$ is bounded below by the CRB, expressed as:

\begin{align}
	&\mathbb{E}\left[(\tilde{o}^k_{rv,t_\tau}-o^k_{rv,t_\tau})(\tilde{o}^k_{rv,t_\tau}-o^k_{rv,t_\tau})^H \right] \succeq \textbf{J}^{-1}(o^k_{rv,t}).
\end{align}

Specifically, the lower bounds for the MSE of $d^k_{rv,t}$ and $\theta^k_{rv,t}$ are represented as:
\begin{align}
	&\mathbb{E}\left[(\tilde{\theta}^k_{rv,t_\tau}-\theta^k_{rv,t_\tau})^2 \right] \geq J_{11} \overset{\Delta}{=} \Psi (\theta^k_{rv,t},f^k_{rv,t}),\\
	&\mathbb{E}\left[(\tilde{d}^k_{rv,t_\tau}-d^k_{rv,t_\tau})^2 \right] \geq J_{22} \overset{\Delta}{=} \Psi (d^k_{rv,t},f^k_{rv,t}),
\end{align}
where $J_{ij}$ represents the $(i, j)$-th element of the inverse of $\textbf{J}^{-1}(o^k_{rv,t})$.

This paper aims to develop a beamformer design that minimizes joint CRBs while adhering to constraints on communication performance and transmit power. To support diverse joint CRB functions within the proposed neural network framework, a utility function is crafted to measure sensing performance, while the achievable sum-rate is employed as a metric to evaluate communication efficiency. As a result, the optimization problem is reformulated as follows:
\begin{subequations}\label{RSU}
\begin{align}
	\min_{\textbf{F}}~&\mathbb{E}\{\Phi(\boldsymbol{\theta},\textbf{F})+\Phi(\boldsymbol{d},\textbf{F})\}\\
	s.t.~&\mathbb{E}\left\lbrace \sum_{k}\sum_{r}R^k_{rv,t_\tau} (\textbf{h}_{rv,t_\tau}, \textbf{F}_{r,t_\tau})\right\rbrace \geq R^{\text{th}}_v,\\
	&\left\|\textbf{F}_r\right\|_F \leq P^{\text{max}}_r,
\end{align}
\end{subequations}
where the expectation $\mathbb{E}(\cdot)$ is computed over the AWGN in the transmission process and the stochastic channel realizations. Furthermore, $R^{\text{th}}_v$ and $P^{\text{max}}_r$ represent the minimum required communication sum-rate and the maximum allowable transmit power, respectively. Solving problem (\ref{RSU}) poses significant challenges due to its variational nature and the non-convexity of the objective function, particularly for traditional model-driven algorithms. To tackle this complexity, we introduce a deep learning-based method designed to efficiently determine the parameters and solve problem (\ref{RSU}).
%All CAVs make their decisions concurrently in a distributed manner, relying on limited knowledge of the entire system. Each CAV has the ability to request RSUs to monitor the states of other vehicles that provide useful information and share this data through V2I communication. The primary objective for each CAV is to minimize the CR ratio by selecting an appropriate subset of external information. Thus, the optimization problem for a specific CAV $v$ is formulated as follows:  
CAVs make distributed decisions with limited system knowledge. Each CAV can request RSUs to share useful external information via V2I, aiming to minimize its collision risk (CR) ratio by selecting an optimal subset of external data. The optimization for a CAV $v$ is thus formulated as:
\begin{subequations}\label{CAV}
	\begin{align}
		\min_{\substack{\boldsymbol{\alpha}_v,\boldsymbol{u}_v,\boldsymbol{\beta}_v,\mathcal{DL}^{\text{th}}_v}}~~&\frac{1}{T}\sum_{\tau}\sum_{l\in\mathcal{L}}\mathbb{E}\{\Xi^{l}_{v,t_\tau}\}\\\label{P-2}
		\text{s.t.}~~~~~&\boldsymbol{\beta}_v\in \{0,1\},(\ref{eq-1})-(\ref{eq-5}), (\ref{eq-16}),
	\end{align}
\end{subequations}

\section{The Proposed Two-Time Scale MA-DDPG Framework}\label{Solution}
The optimization problems in (\ref{RSU}) and (\ref{CAV}) are highly nonconvex and heavily constrained, and their solution is further complicated by time-varying CSI; classical static optimization methods therefore either produce suboptimal outcomes or require costly per-slot recalibration and cannot scale to multi-stage setups that span many coherence intervals. To overcome these limitations, we recast request design and resource allocation as a Multi-Agent Reinforcement Learning problem in which each CAV and the BS act as autonomous agents that make decentralized decisions from local observations; this formulation removes the need for repeated global optimization, enables online adaptation to evolving radio conditions, and permits agents to coordinate resource usage while minimizing collision risk.

A Markov model can be employed to represent the movement parameter control and resource allocation challenges, with the BS and each CAV serving as distinct agents. Notably, the optimization problem defined in (\ref{CAV}) is addressed for both short-term and long-term horizons, while the problem in (\ref{RSU}) is resolved exclusively for long-term scenarios. At a given time $t\in\{t_{\tau},\tau\}$, each agent $j\in\{\mathcal{R},\mathcal{V}\}$ derives a portion of the global state, referred to as its local observation $\tilde{s}_{j,t}$. Following this, the agent selects an action $a_{j,t}$, receives a reward $r_{j,t}$, and the environment transitions to the subsequent state $\tilde{s}_{j,(t+1)}$.

\subsubsection{State Space and Local Observation}
The global state, $\tilde{s}_{t}$, includes all the environment components, e.g., channel conditions, movement parameters, and the agent's
behavior. However, only a subset of the global state can be observed by each agent. The local observation of
each agent $j$ includes channel information such as channel gain as well as movement parameter obtained by sensors of CAV $v$. 

\subsubsection{Action Space}
The action space for CAV's agents includes exogenous information request, and movement parameter determination. 
\subsubsection{Reward Function}
Our aim is to maximize the utility functions defined in (\ref{U-v}) for the proposed resource allocation problem. Therefore, the instant reward function for agent $j$ is defined as
\begin{align}\label{U-v}
	&r_{j,t}=\begin{cases}
		\frac{1}{T}\sum_{\tau}\sum_{l\in\mathcal{L}}\mathbb{E}\{\Xi^{l}_{v,t_\tau}\}, & \text{$j=v$}.\\
		\mathbb{E}\{\Phi(\boldsymbol{\theta},\textbf{F})+\Phi(\boldsymbol{d},\textbf{F})\}, & \text{$j=r$}.
	\end{cases}
\end{align}

%The ultimate goal is to increase the reward of the whole network, which is expressed in the learning section as the total reward of the base station and the vehicles as $r_t=\mathfrak{U}_{BS,t}-\sum_{v}\mathfrak{U}_{v,t}$.

\subsubsection{Reinforcement Learning}
In the MARL problem, the actions of each agent are selected under its policy, $\tilde{\boldsymbol{\pi}}_j:\Lambda(\mathcal{A}_j)\leftarrow\mathcal{S}$, where $\Lambda$ denotes a probability distribution. Moreover, in our resource allocation problem, the joint policy $\tilde{\boldsymbol{\pi}}$ is defined as $\tilde{\boldsymbol{\pi}}=\{\tilde{\boldsymbol{\pi}}_j\}_{j\in\{\mathcal{V},\mathcal{R}\}}$. A policy $\tilde{\boldsymbol{\pi}}_{j}=[\tilde{\mu}_{j,0},\dots,\tilde{\mu}_{j,t},\dots,\tilde{\mu}_{j,T-1}]$ comprises of $T$ functions $\tilde{\mu}_{j,t}$, where $a_{j,t}=\tilde{\mu}_{j,t}(\tilde{s}_{j,t})$ for each time step $t$.

The aim is to determine the optimal policy $\tilde{\pi}^*_j$ that maximizes the expected total reward $\tilde{\pi}^*_j=\arg\max_{\tilde{\pi}_j}\tilde{J}_{\tilde{\pi}_j}$, where $\tilde{J}^*=\mathbb{E}_{\tilde{s}_{j,t}}\{\tilde{V}^*_{j,t}(\tilde{s}_{j,t})\}$ is the episode expected reward. The state value function is formulated as $\tilde{V}_{j,t}(\tilde{s}_{j,t})=\mathbb{E}_{\tilde{\pi}_j}[G_{j,t}|\tilde{s}_{j,t}=\tilde{s}]$, where $G_{j,t}=\sum_{i}^{\infty}\gamma^{i}r_{j,t+i}$ and $\gamma$ denotes the discount factor to balance the instant and future rewards. The value functions satisfy the Bellman
equation, and thus can be expressed as
\begin{align}\label{V-Value}
	\tilde{V}_{j,t}(\tilde{s}_{j,t})=\mathbb{E}_{\tilde{\pi}_j}[r_{j,t+1}+\gamma \tilde{V}_{j,t+1}(\tilde{s}_{j,t+1})|\tilde{s}_{j,t}=\tilde{s}].
\end{align}
Based on the value function in the value based RL problem, the Q-value that represents the value from state $\tilde{s}_{j,t}$ and action $a_{j,t}$ over policy $\pi_j$, is described as
\begin{align}\label{Q-Value}
	\tilde{Q}_{j,t}(\tilde{s}_{j,t},a_{j,t})=r_{j,t}(\tilde{s}_{j,t},a_{j,t})+\gamma\mathbb{E}_{\tilde{\pi}_j}[\tilde{V}_{j,t}(\tilde{s}_{j,(t+1)})].
\end{align}

The goal is to find the optimal policy $\tilde{\pi}^*_j$ that maximize the value function. 
Since our movement parameter control and resource allocation problem is high dimensional with continuous action space, we utilize multi-agent deep
deterministic policy gradient (MA-DDPG) framework, in which both the Q-value and policy are modeled as neural networks. Although MA-DDPG requires the global information in the training phase, it can be executed distributively in the execution phase by each agent, where each agent has one actor and one critic network. The details of the proposed steps  is shown in Algorithm I. 
The MADDPG agent $j$ employs two	main deep neural networks: actor network with $\theta^{\mu}_j$ parameter to approximate the deterministic policy $\mu_j(s_j|\theta^{\mu}_j)$ and a critic network with $\theta^{Q}$ parameter to approximate a state-value function $Q_j(s_j,a_j|\theta^Q_j)$. In order to train the main networks, a random mini-batch consisting of $\bar{D}_j$ samples $(s^i_j, a^i, r^i_j, s^{i+1}_j)|_{i=1}^{\bar{D}_j}$ is selected from the replay buffer $\bar{D}_j$. Each sample is assigned an index denoted by $i$. The parameters $\theta^{Q}$ and $\theta^{\mu}$ of each critic and actor networks are updated as
\begin{align}\label{Loss-Function}
	\mathcal{L}_j(\theta^Q_j)=\frac{1}{\bar{D}_j}\sum_i(y^i_j-Q_j(s^i_j,a^i|\theta^Q_j))^2,
\end{align}
\begin{align}\label{Policy}
	\nabla_{\theta^{\mu}_j}J=\frac{1}{\bar{D}_j}\sum_i\nabla_{a_j}Q_{j}(s^i_j,a^i|\theta^Q_j)\nabla_{\theta^{\mu}_j}\mu_j(s^i_j|\theta^{\mu}_j),
\end{align}
where $y^i_j=r^i_j+\gamma Q'_j(s^{i+1}_j,a^{i+1}|\theta^{Q'}_j)$ denotes the target value. The actor and critic networks are softly updated for the target parameters $\theta^{\mu'}_j$ and $\theta^{Q'}_j$ as follows:
\begin{align}\label{eq-mu}
	\theta^{\mu'}_j \leftarrow \varsigma_j \theta^{\mu}_j+(1-\varsigma_j) \theta^{\mu'}_j,
\end{align}
\begin{align}\label{eq-Q}
	\theta^{Q'}_j\leftarrow \varsigma_j \theta^{Q}_j+(1-\varsigma_j) \theta^{Q'}_j.
\end{align}
where $\varsigma_j$ is a constant close to zero.
\begin{algorithm}[t!]
	\caption{\small{Two-Time Scale MA-DDPG Framework}}
	\fontsize{8}{8}\selectfont
	\label{Refined_MADDPG}
	\begin{algorithmic}[1]
		\State \textbf{Initialize:} For each agent $j \in \{\mathcal{V}, \mathcal{R}\}$:
		\State \quad Initialize actor $\mu_j$, critic $Q_j$, and their target networks.
		\State \quad Initialize replay buffer $\bar{D}_j$ and set learning rates $\varsigma_j$.
		\For{each episode}
		\State Initialize short-term state $s_0^{S}$ and long-term state $s_0^{L}$.
		\For{each long-term step $\tau$}
		\State Each agent observes $s_{j,\tau}^{L}$ and selects long-term action $a_{j,\tau}^{L}$.
		\State Execute $a_{j,\tau}^{L}$ (e.g., acceleration, heading) and receive $r_{j,\tau}^{L}$.
		\State Store $(s_{j,\tau}^{L}, a_{j,\tau}^{L}, r_{j,\tau}^{L}, s_{j,\tau+1}^{L})$ in $\bar{D}_j$.
		
		\For{each short-term step $t_\tau$}
		\State Observe $s_{j,t_\tau}^{S}$ and select $a_{j,t_\tau}^{S}$ using $\mu_j$ with exploration noise.
		\State Execute action and receive reward $r_{j,t_\tau}^{S}$ and next state $s_{j,t_\tau+1}^{S}$.
		\State Store $(s_{j,t_\tau}^{S}, a_{j,t_\tau}^{S}, r_{j,t_\tau}^{S}, s_{j,t_\tau+1}^{S})$ in $\bar{D}_j$.
		\If{update condition met}
		\State Sample mini-batch from $\bar{D}_j$.
		\State Update critic $Q_j$ via loss \eqref{Loss-Function}.
		\State Update actor $\mu_j$ using policy gradient \eqref{Policy}.
		\State Soft update target networks via \eqref{eq-mu}, \eqref{eq-Q}.
		\EndIf
		\EndFor
		\EndFor
		\If{episode mod update\_interval = 0}
		\State Repeat actor/critic updates using long-term memory $\tilde{D}_j$.
		\EndIf
		\EndFor
	\end{algorithmic}
\end{algorithm}
\section{Computational Complexity Analysis}\label{Signaling Overhead} 
DDPG mainly includes a replay buffer and four neural networks. Assuming that the actor network contains $\bar{L}$ fully connected layers and the critic network contains $\bar{K}$ fully connected layers. Thus the time complexity and space complexity of DDPG can be derived with regard to floating point operations per second (FLOPS) \cite{Qiu2019}.

The neural networks for every layer have a vector $\bar{n}^{\text{Act}}_{l}$ and a matrix $\bar{n}^{\text{Act}}_{l} \times \bar{n}^{\text{Act}}_{l+1}$ for a fully connected layer to perform dot product. The FLOPS computation is $(2\bar{n}^{\text{Act}}_{l}-1) \times \bar{n}^{\text{Act}}_{l+1}$, i.e., multiply $\bar{n}^{\text{Act}}_{l}$ times and add $\bar{n}^{\text{Act}}_{l}-1$ times. Activation layers also should be taken into consideration, which is calculated without dot product. It is only measured by FLOPS, where addition, subtraction, multiplication, division, exponent, and square root are counted as a single FLOPS. So the time complexity can be defined as follows:
	\begin{align}
		\nonumber& 2\sum_{l=0}^{\bar{L}-1} ((2\bar{n}^{\text{Act}}_{l}-1)  \bar{n}^{\text{Act}}_{l+1} + \nu_{\text{A}} \bar{n}^{\text{Act}}_{l+1}) 
		+ 2\sum_{k=0}^{\bar{K}-1} ((2\bar{n}^{\text{Cri}}_{k}-1) \bar{n}^{\text{Cri}}_{k+1} \\&+ \nu_{\text{A}} \bar{n}^{\text{Cri}}_{k+1}) 
		 = \mathcal{O}(\sum_{l=0}^{\bar{L}-1} \bar{n}^{\text{Act}}_{l}\bar{n}^{\text{Act}}_{l+1} + \sum_{k=0}^{\bar{K}-1} \bar{n}^{\text{Cri}}_{k}\bar{n}^{\text{Cri}}_{k+1})  = \mathcal{O}({n}^2)
	\end{align}
	where $\nu_{\text{A}}$ is the corresponding parameter of the activation layer \cite{Qiu2019}.

\section{Simulation Results}\label{Simulation}
%\subsection{The Setup}
In this section, we evaluate the performance of our proposed approach. We consider a transportation system with $V=100$ CAVs and a communication system with one BS. The main simulation parameters including the technical constraints, the parameters used for training the actor and critic networks, and operational parameters are summarized in Table \ref{simulation_params} \cite{vu2020cell}. 
We assume that  small-scale and large-scale fading parameters are updated every coherence time, and every 100 coherence times, respectively. The control environment and MA-DDPG algorithms are implemented in Python using Tensorflow 1.14. 
\begin{table}[h]
	\caption{Simulation parameters}\label{simulation_params} \centering
	\footnotesize
	\fontsize{7}{7}\selectfont
	\begin{tabular}{p{2.9cm} p{.6cm} p{2.9cm} p{.6cm} }
		\hline
		\textbf{Parameter} & \textbf{Value} & \textbf{Parameter} & \textbf{Value}\\
		\hline\hline
		Number of RSUs & 4  &
		Maximum transmit power of RSU & 23 dBm \\
		Noise power & -114 dBm\\
		Number of antennas at RSU & 32 &
		Height of RSU & 15 m\\
		Length of vehicle & 4 m &
		Maximum acceleration of vehicle & 5 m/s$^2$ \\
		Minimum acceleration of vehicle & -3 m/s$^2$ &
		Maximum velocity of vehicle & 40 m/s \\
		&  &
		Number of vehicles  & 300 \\
		Number of lanes & 4 &
		Lane width & 4 m \\
		Actuation lag & 20 ms &
		&  \\
		Duration of actuation lag & 20 ms&
		BSM size & 1000 bytes\\
		Bandwidth of C-V2X & 20 MHz   &
		&  \\
		Learning rate of critics network & 0.0001 &
		Discount factor for large/short-term time & 0.95, 0 \\
		Target network update frequency & 1000 &
		Number of hidden layers & 2\\
		Number of neurons for layer & 512 &
		Replay buffer size: long/short-term time & 10$^4$, 10$^6$ \\
		Minibatch size & 64 &
		Number of neurons in layer & 256 \\
		Activation function & ReLU &
		Optimizer for DNNs & Adam \\
		\hline
	\end{tabular}
\end{table} 

In order to evaluate the effectiveness of our proposed scheme, we compare it with the following existing benchmark schemes:
\begin{itemize}
	\item \textbf{Benchmark 1}: 
%	The work presented in \cite{dou2023sensing}, where the authors proposed a sensing scheduling paradigm of an ISAC system, in which the ISAC base station simultaneously provides data transmissions to a group of users via non-orthogonal multiple access. To measure the performance, they adopted the sensing estimation mutual information and guarantee the ISAC base station should extract a required amount of mutual information from the echoes of sensing targets which are scheduled to be sensed. They have proposed a joint optimization of the beamforming, the transmission duration and the sensing scheduling, with the objective of maximizing the sensing efficiency (i.e., the number of the selected sensing targets over the transmission duration).
In \cite{dou2023sensing}, the authors proposed a sensing scheduling paradigm for ISAC systems, where the base station simultaneously serves multiple users via non-orthogonal multiple access. System performance was evaluated using sensing estimation mutual information, ensuring that the base station extracts the required information from scheduled sensing targets. To this end, they formulated a joint optimization of beamforming, transmission duration, and sensing scheduling, with the objective of maximizing sensing efficiency, defined as the number of selected sensing targets per transmission duration.
	\item \textbf{Benchmark 2}: 
%	The work presented in \cite{liu2023toward}, where the authors studied a reconfigurable intelligent surface empowered ISAC network in which the base station simultaneously executes both sensing and communication by sending the combined information-bearing and dedicated sensing waveforms in the entire wireless space. They jointly optimized the transmit beamforming at the base station as well as the transmission and reflection beamforming to maximize the sensing SNR.
In \cite{liu2023toward}, the authors investigated a RIS-empowered ISAC network, where the base station performs joint sensing and communication by transmitting both information-bearing and dedicated sensing waveforms across the wireless medium. They formulated a joint optimization of the base station’s transmit beamforming and the RIS reflection beamforming, aiming to maximize the sensing SNR.
\end{itemize}

\subsection{Training and Convergence Properties}
\begin{figure*}[!t]
	\centering
	\subfigure[]{\includegraphics[width=0.3\textwidth]{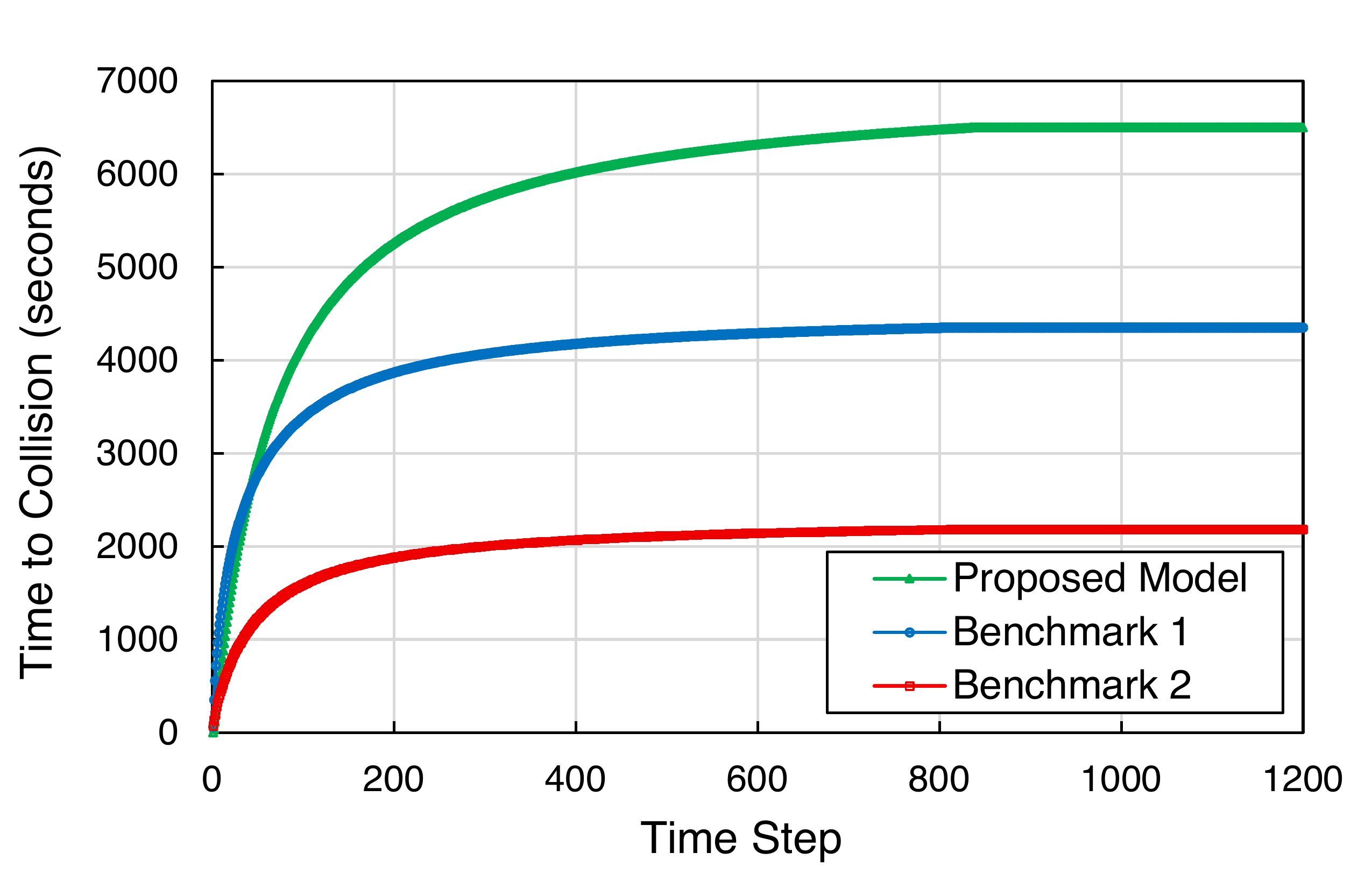}}
	\subfigure[]{\includegraphics[width=0.3\textwidth]{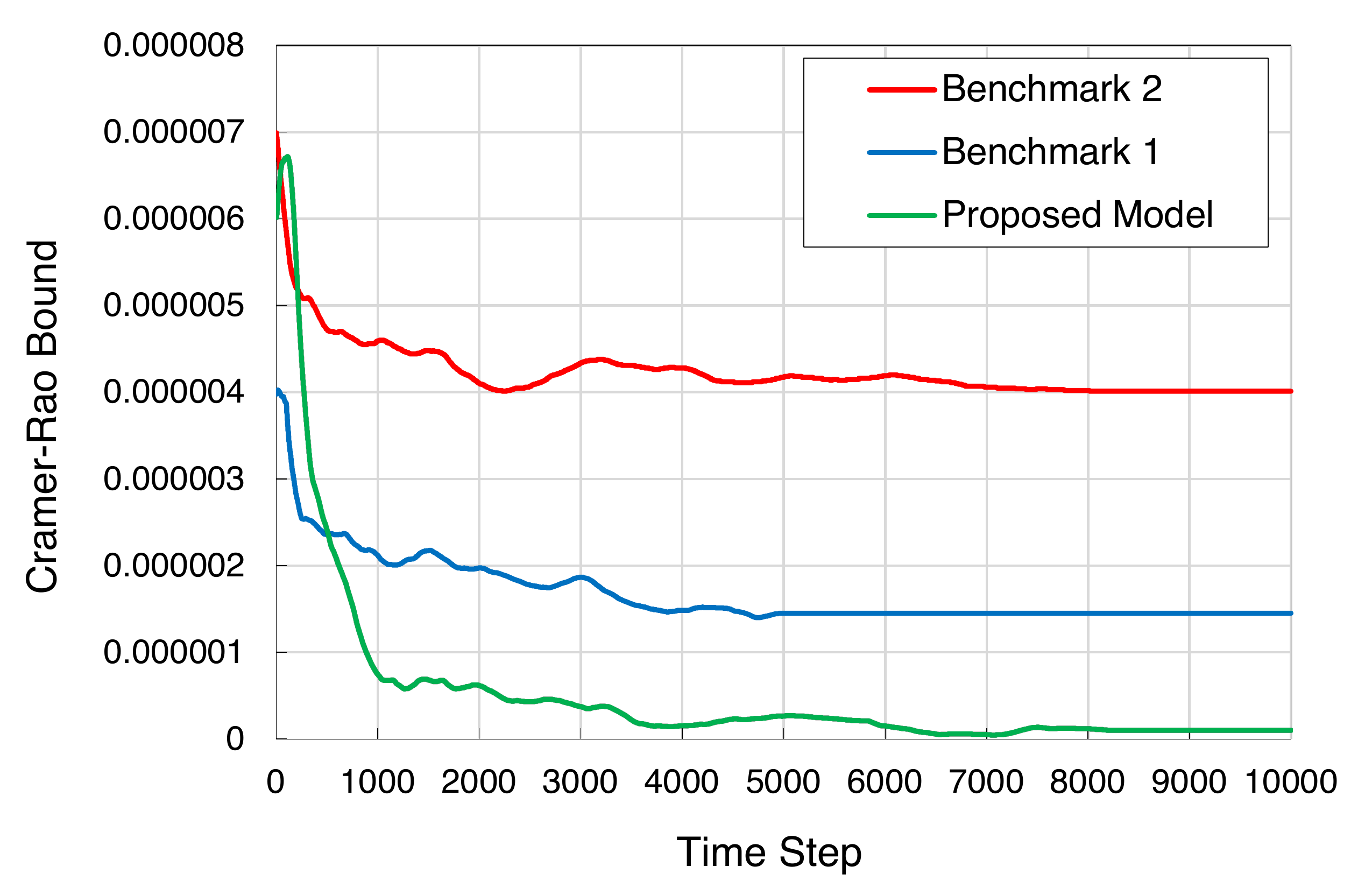}}
	\subfigure[]{\includegraphics[width=0.3\textwidth]{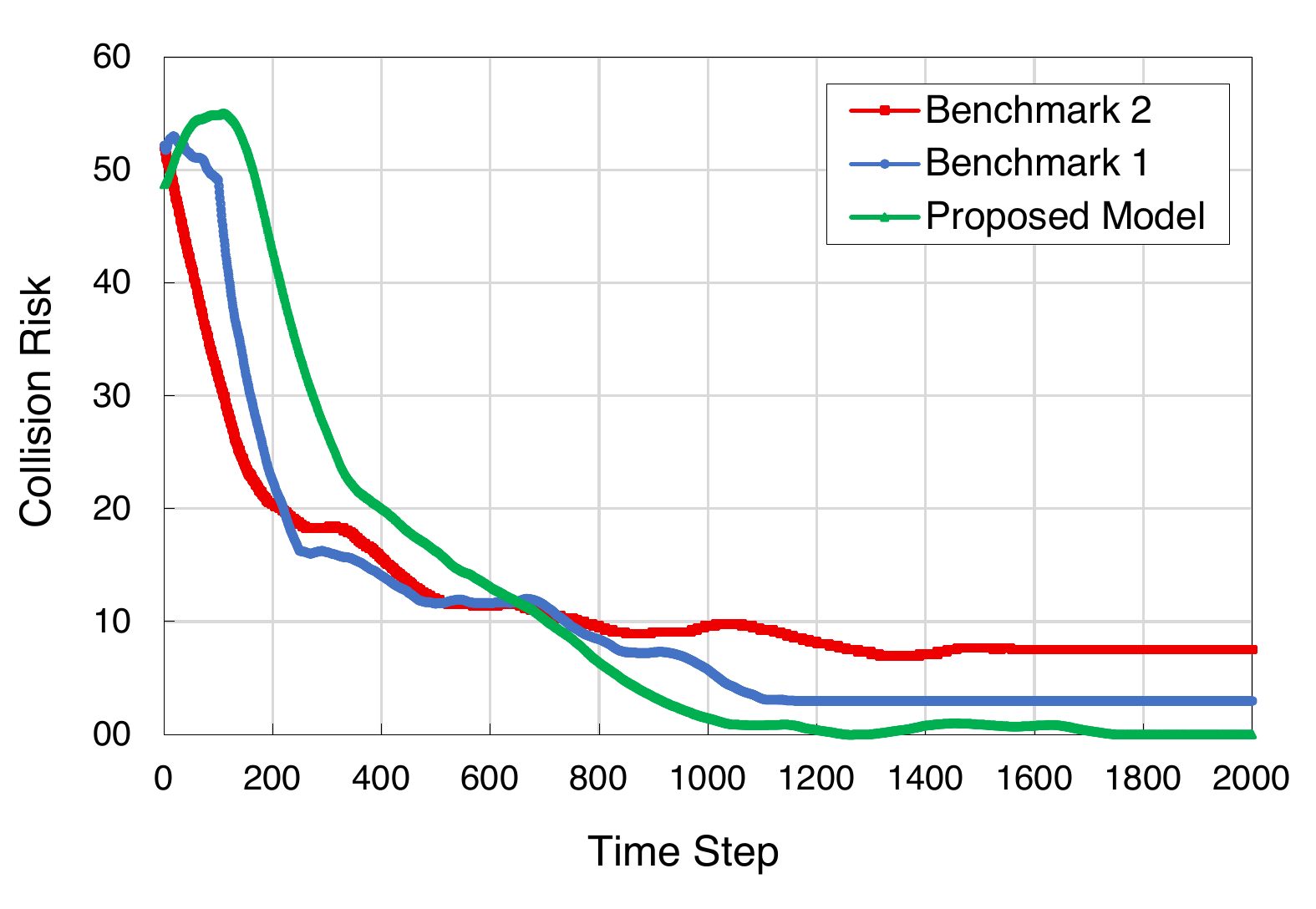}}
	\caption{Convergence of the (a) TTC and (b) CRB, (c) CR over time}
	\label{one}
\end{figure*}

%We analyze the convergence of our algorithm alongside the models from \cite{dou2023sensing} and \cite{liu2023toward} based on their TTCs and CRB performance. We see the convergence of TTCs for CAVs in Fig. \ref{one}(a) and also that of CRBs for RSUs in Fig. \ref{one}(b), which both validates the effectiveness of our proposed algorithm. As shown in the figure, our proposed algorithm achieves the best performance in terms of the highest TTCs and the lowest CRBs. The reasons can be explained as follows. The models in \cite{dou2023sensing} and \cite{liu2023toward}, which both do not consider Doppler spread for vehicles, result in lower TTCs and higher CRBs. 
%The model in \cite{dou2023sensing}, which exploits the advanced successive interference cancellation in a non-orthogonal multiple access system, demonstrates better performance because it can enable multiple users’ simultaneous transmissions over the same spectrum channel while effectively mitigating the co-channel interference. Compared to \cite{dou2023sensing} and \cite{liu2023toward} respectively, our proposed model increases TTCs by 33\% and 66\%, as well as decreases CRBs by 32\% and 75\%.
We evaluate the convergence of our algorithm against \cite{dou2023sensing} and \cite{liu2023toward} in terms of TTC for CAVs (Fig. \ref{one}(a)) and CRB for RSUs (Fig. \ref{one}(b)), confirming the effectiveness of our approach. Our algorithm achieves the highest TTC and lowest CRB, as it explicitly accounts for vehicle Doppler spread. In contrast, both baselines neglect Doppler effects, leading to degraded performance. The model in \cite{dou2023sensing} outperforms \cite{liu2023toward} due to its use of successive interference cancellation in NOMA, which supports simultaneous transmissions while mitigating co-channel interference. Overall, our model improves TTC by 33\% and 66\%, and reduces CRB by 32\% and 75\% compared to \cite{dou2023sensing} and \cite{liu2023toward}, respectively.

%Fig. \ref{one}(c) demonstrates the CR over time. The model presented in \cite{liu2023toward} initially reduces the risk of collision very fast, but it does not have good performance in the end. The CR in the model \cite{dou2023sensing} decreases after some iterations and reaches a lower value compared to the previous one. Our proposed model takes time to learn and adjust to the evolving communication environment. It has higher risks at first but no risks at the end. The proposed model outperforms the baselines in terms of reducing CR over time.
Fig. \ref{one}(c) shows the CR evolution over time. While \cite{liu2023toward} reduces CR rapidly at the beginning but converges poorly, and \cite{dou2023sensing} achieves lower CR after several iterations, our proposed model initially exhibits higher risk but progressively adapts to the dynamic environment, ultimately eliminating collisions. Overall, it achieves superior long-term CR reduction compared to both baselines.

\subsection{Scalability of the Proposed Scheme}
\begin{figure*}[!t]
	\centering
	\subfigure[]{\includegraphics[width=0.3\textwidth]{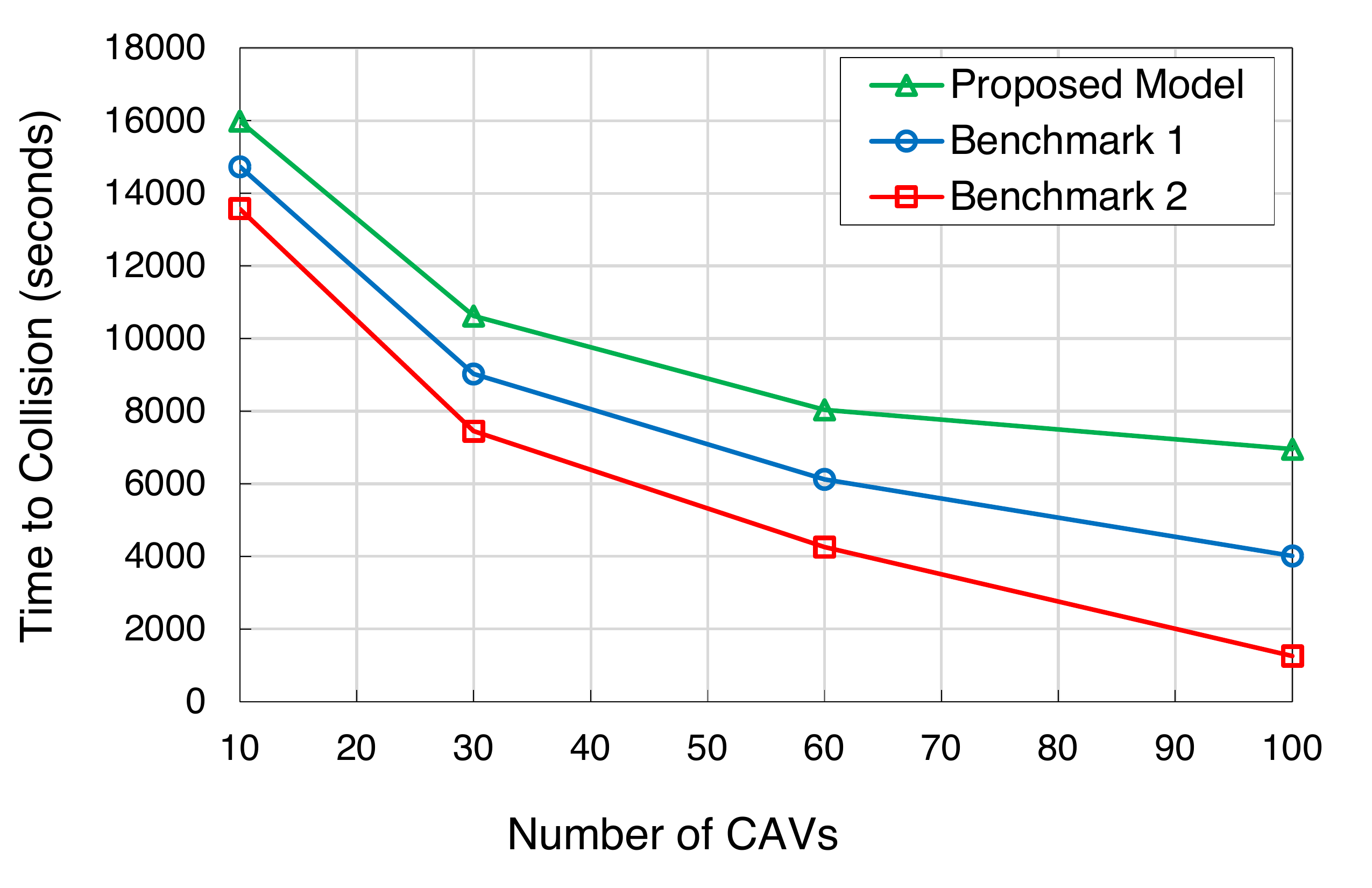}}
	\vspace{-.2cm}
	\subfigure[]{\includegraphics[width=0.3\textwidth]{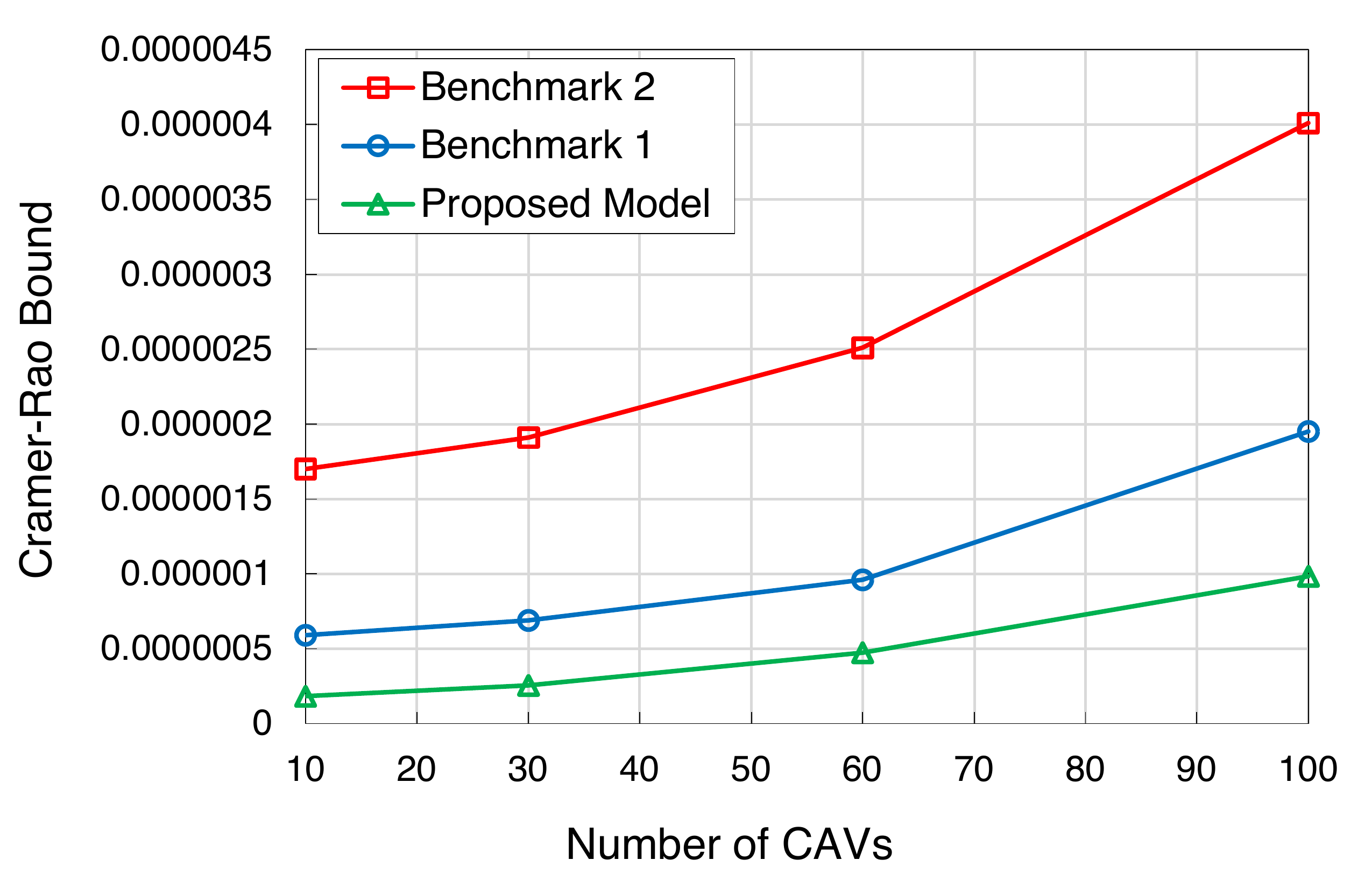}}
	\subfigure[]{\includegraphics[width=0.3\textwidth]{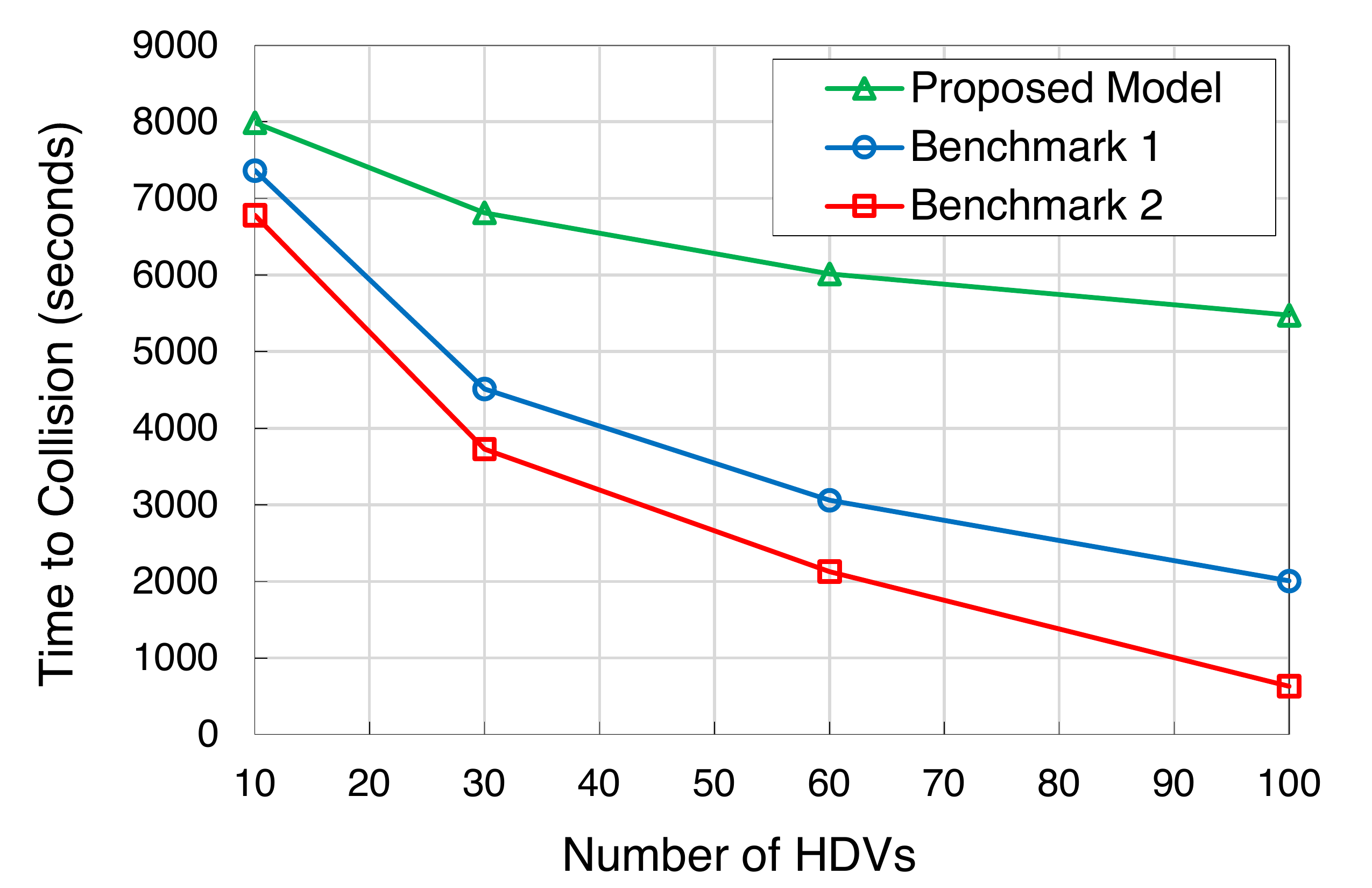}}
	\subfigure[]{\includegraphics[width=0.3\textwidth]{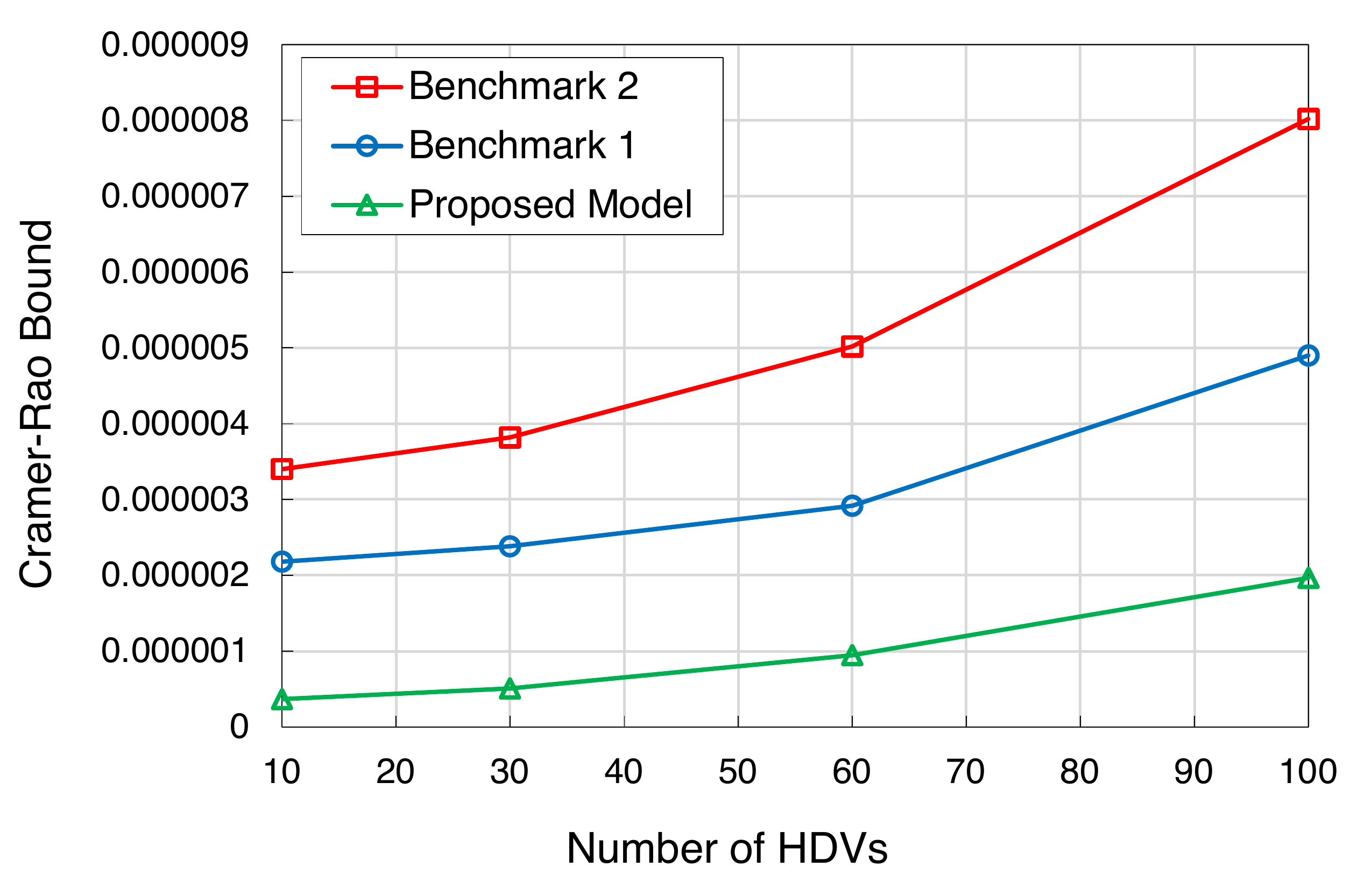}}
	\subfigure[]{\includegraphics[width=0.3\textwidth]{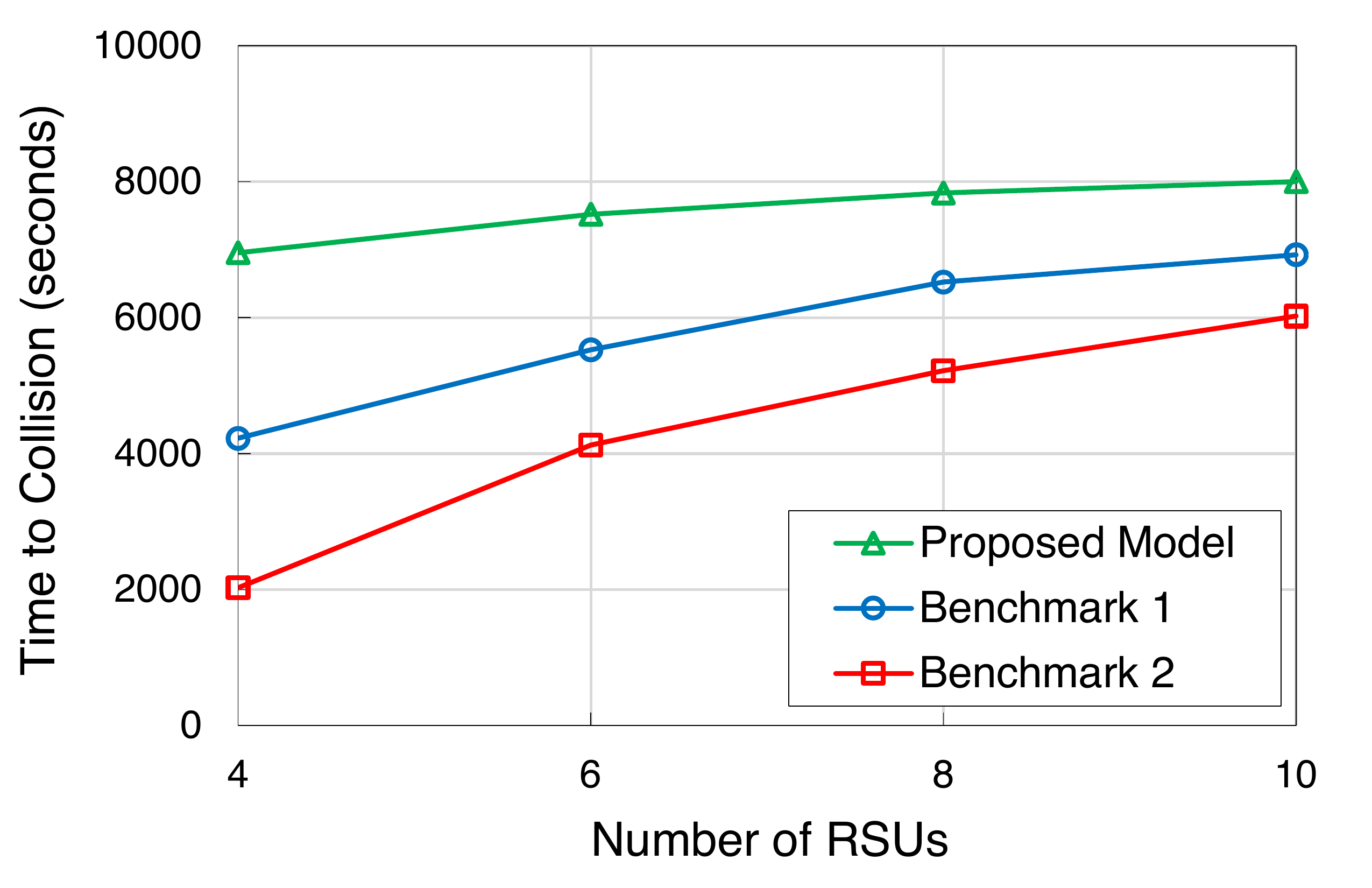}}
	\subfigure[]{\includegraphics[width=0.3\textwidth]{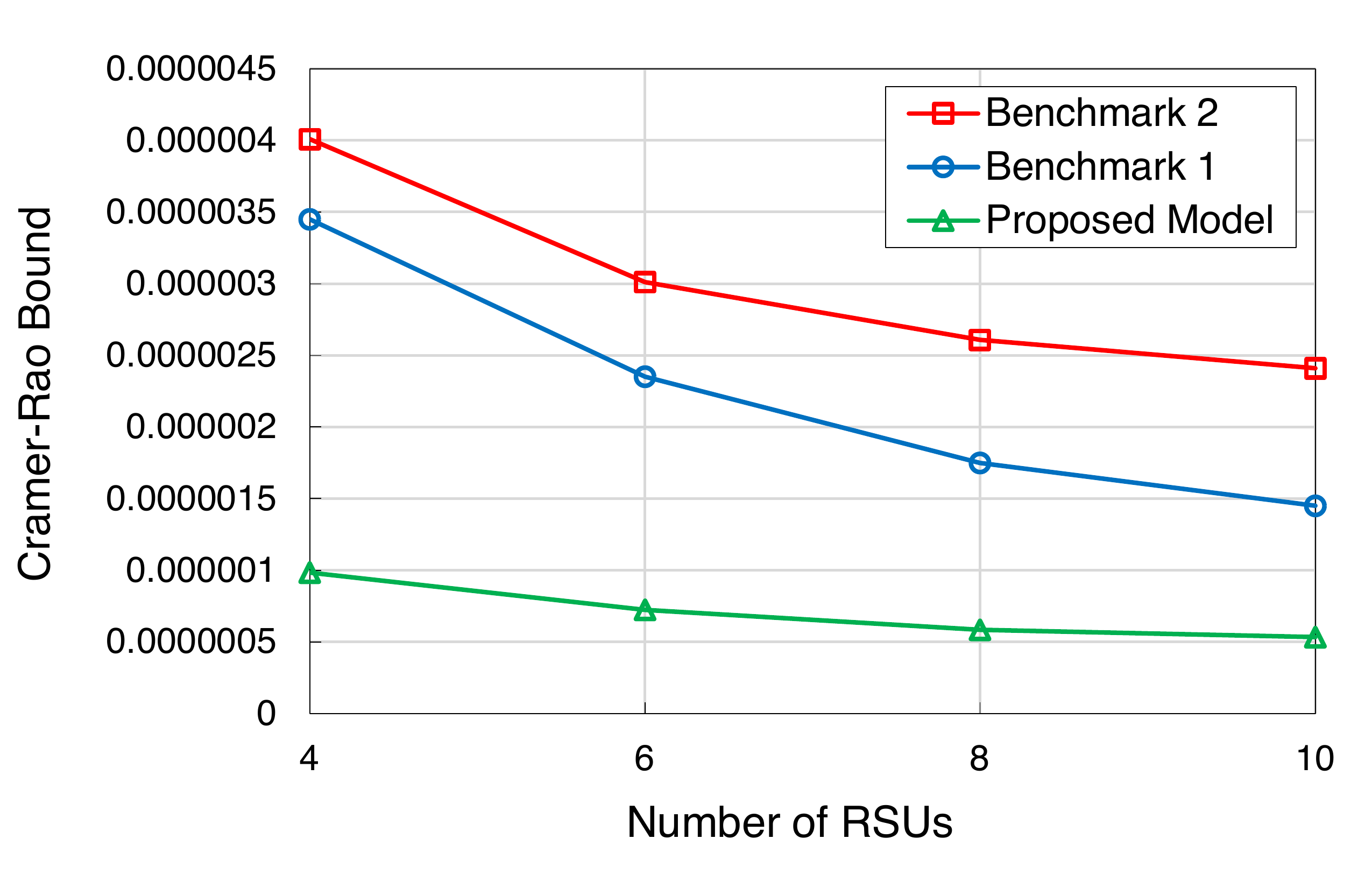}}
	\caption{(a) TTC and (b) CRB under different numbers of CAVs, (c) TTC and (d) CRB under different numbers of HDVs, (e) TTC and (f) CRB with different numbers of RSUs}
	\label{two}
\end{figure*}

\begin{figure*}[!t]
	\centering
	\subfigure[]{\includegraphics[width=0.3\textwidth]{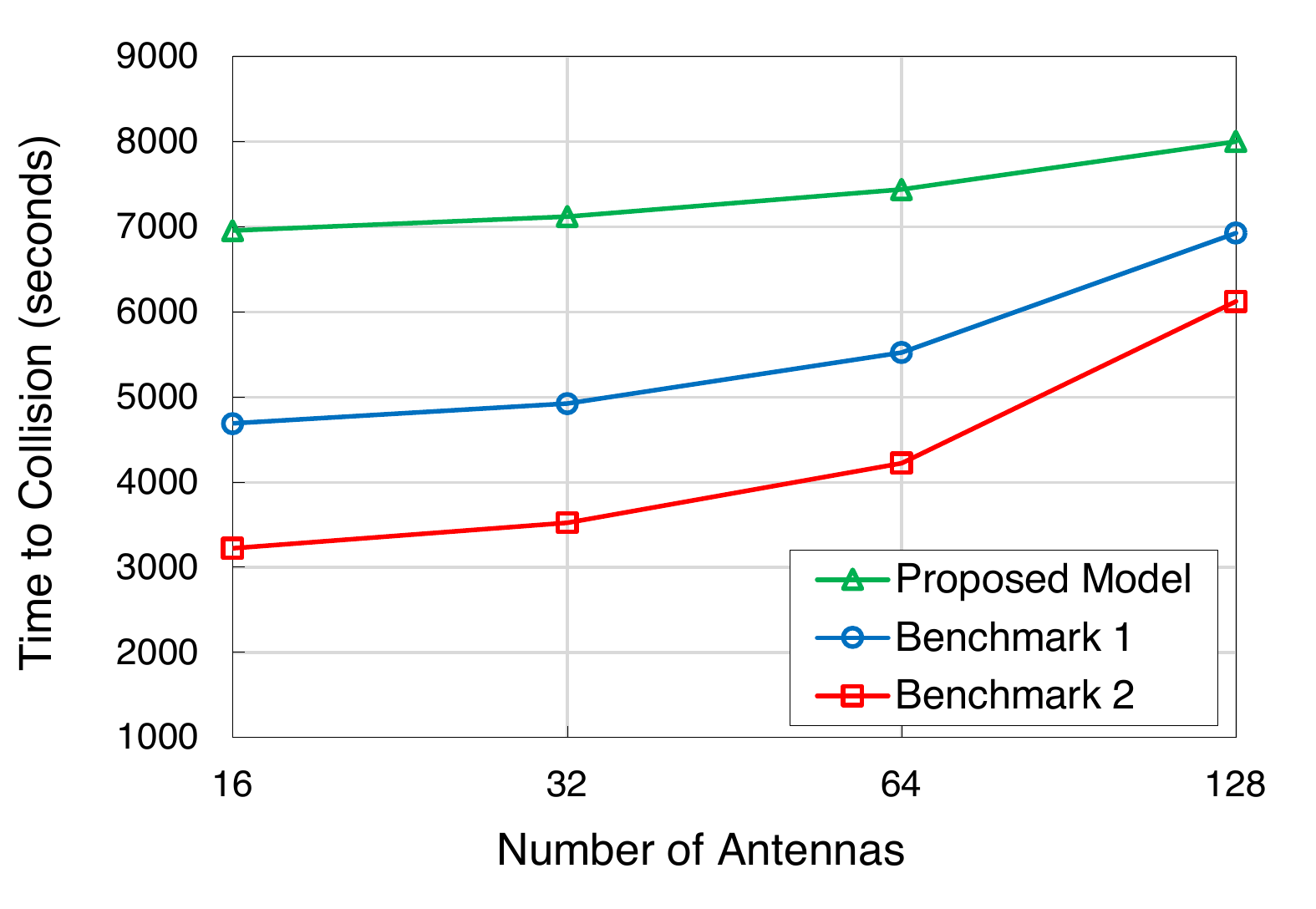}}\vspace{-.4cm}
	\subfigure[]{\includegraphics[width=0.3\textwidth]{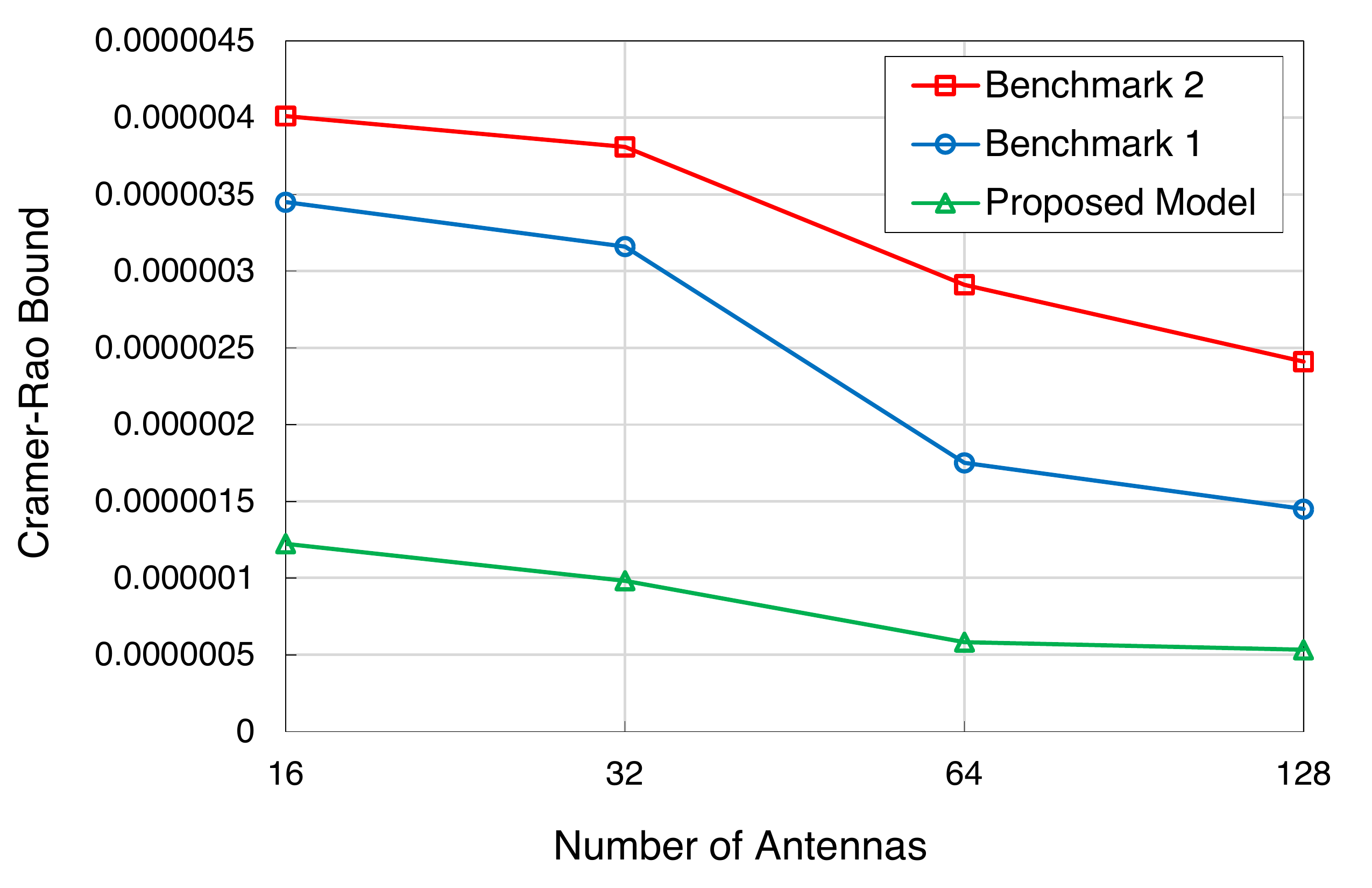}}
	\subfigure[]{\includegraphics[width=0.3\textwidth]{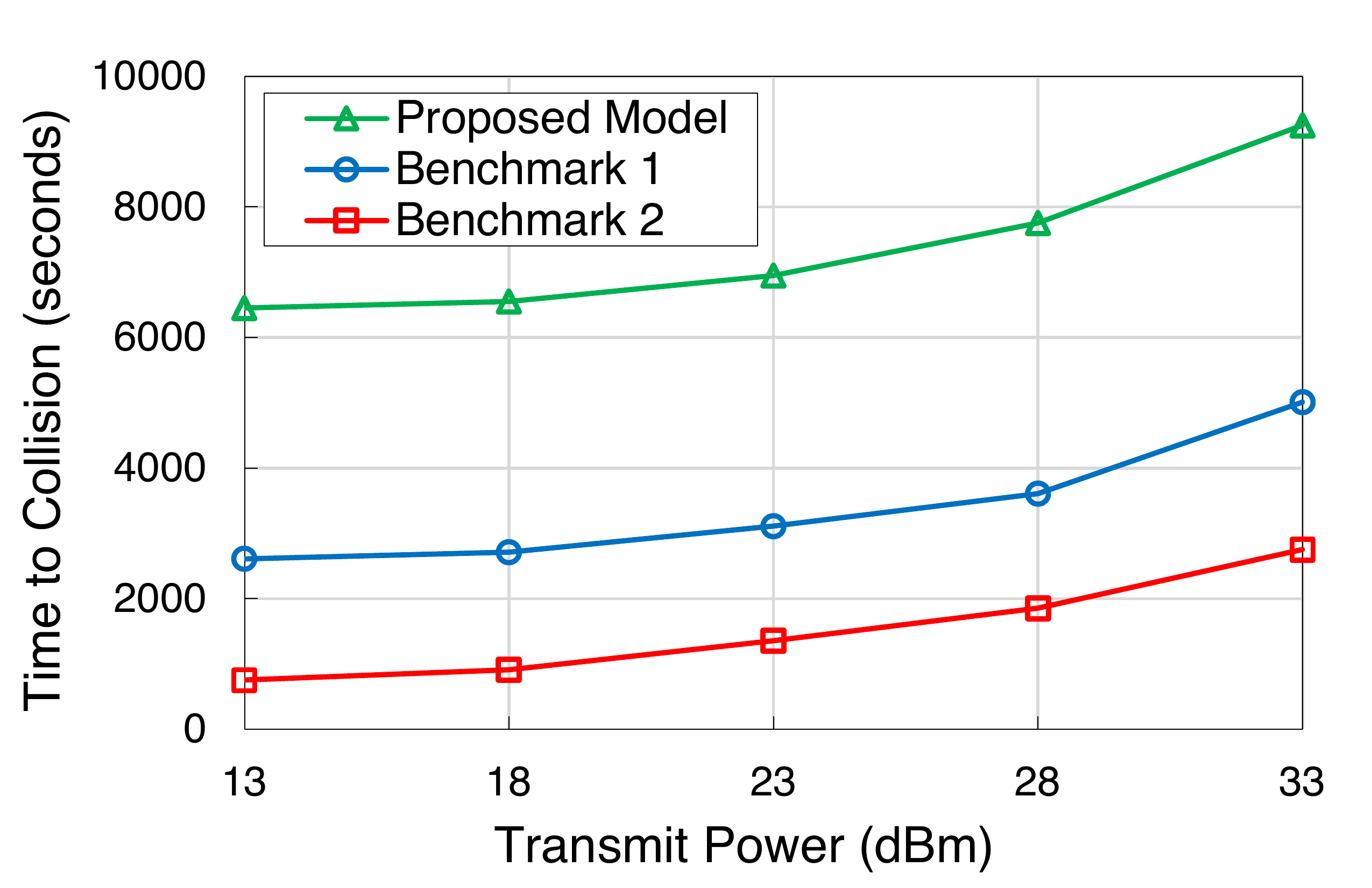}}
	\subfigure[]{\includegraphics[width=0.3\textwidth]{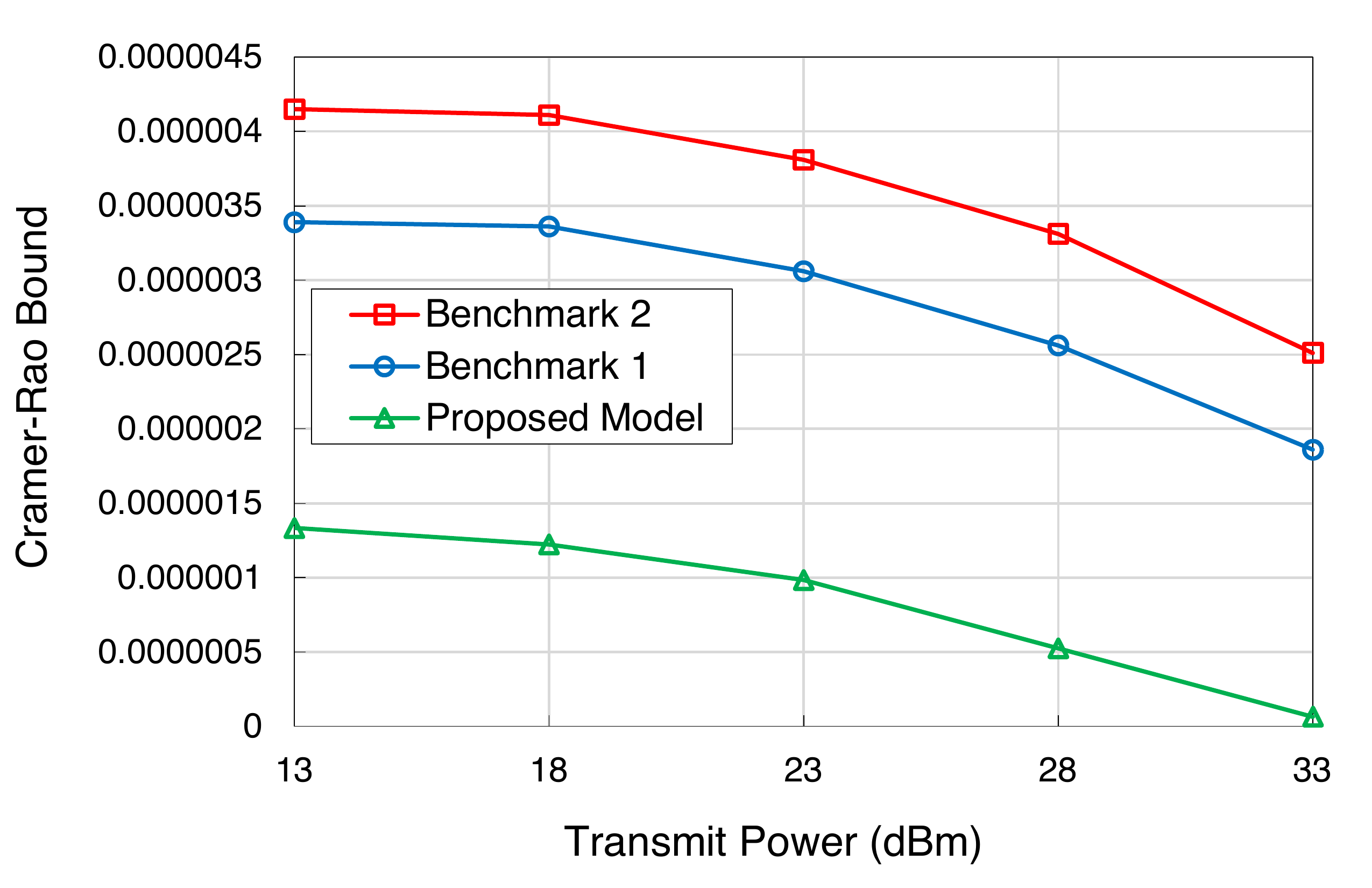}}
	\subfigure[]{\includegraphics[width=0.3\textwidth]{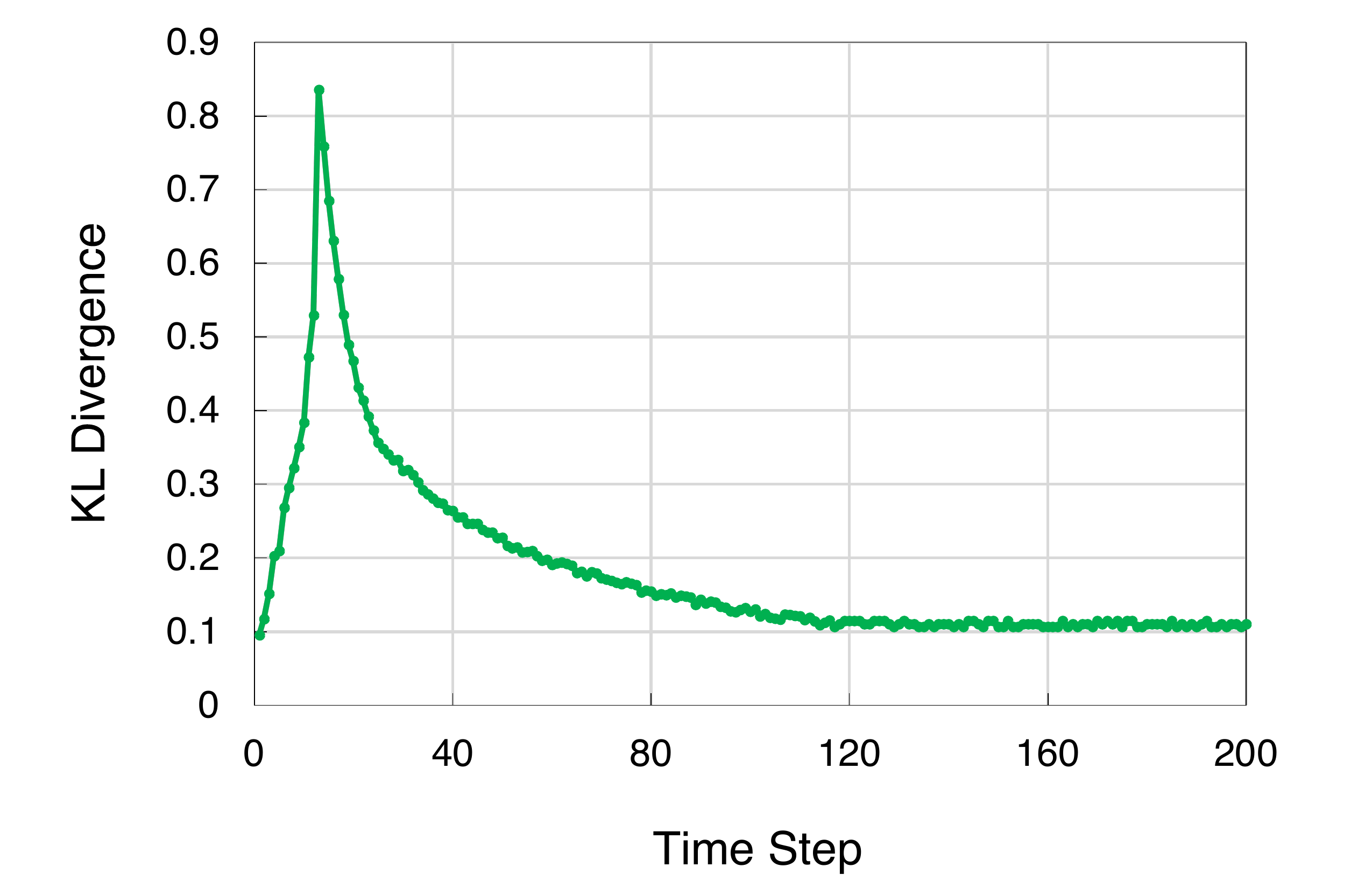}}
	\subfigure[]{\includegraphics[width=0.3\textwidth]{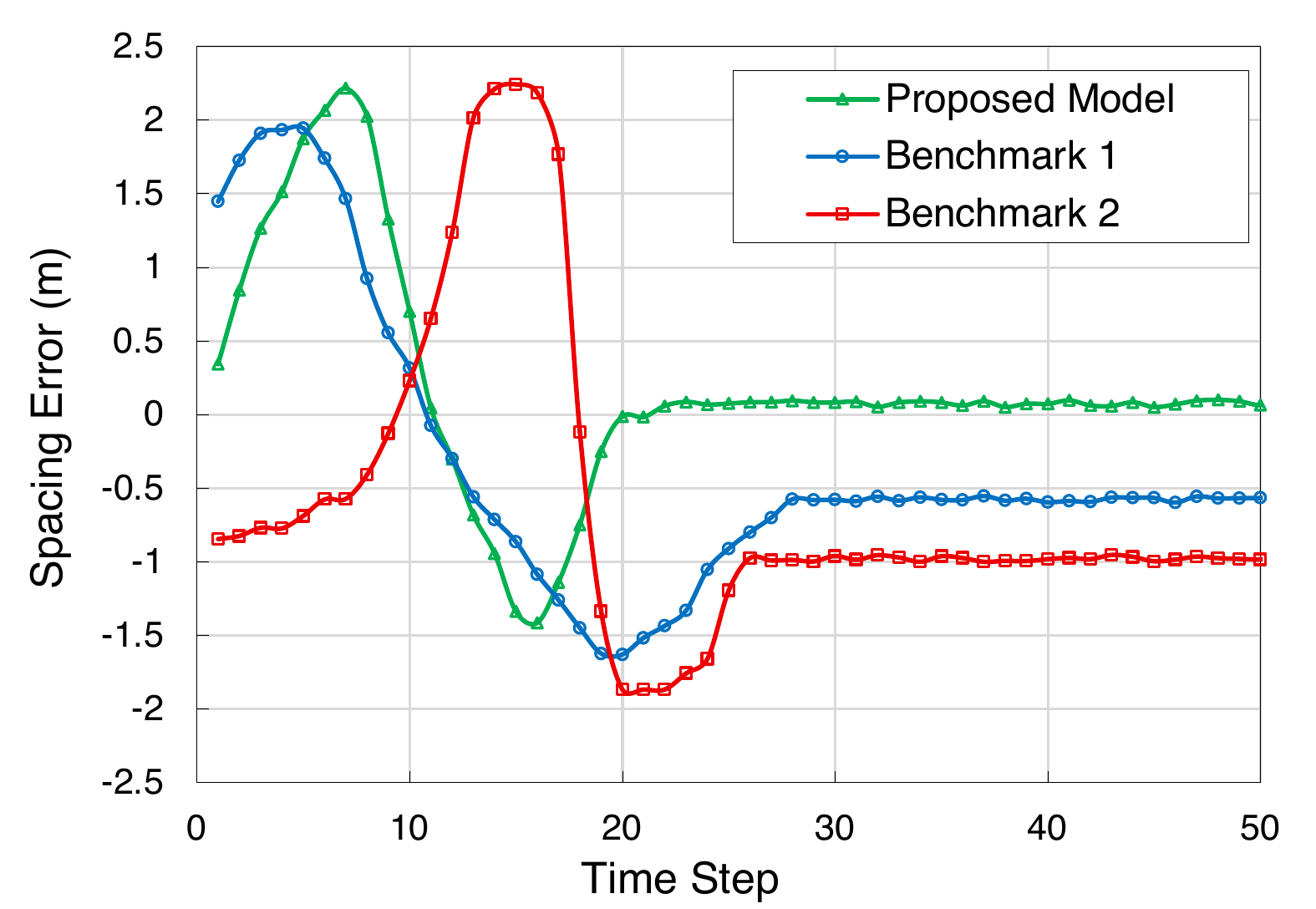}}
	\caption{(a) TTC and (b) CRB with different numbers of antennas, (c) TTC and (d) CRB with different transmit power budget, (e) The KL divergence for each time step, (f) spacing error over time}
	\label{three}
\end{figure*}

In the following, we reveal the influence of the number of vehicles and the number of RSUs on performance and show the result curves in Fig. \ref{two}. In Fig. \ref{two}(a), it can be seen that with increase in the number of CAVs TTC decreases, because more CAVs result in less distance between them which reduces the time to collision. On the other hand, with more CAVs the accuracy of parameter estimation by RSUs decreases, so this is why we see an upward trend for CRB in Fig. \ref{two}(b). Our proposed model shows 22\% and 43\% increase in TTCs as well as 6\% and 8\% decrease in CRBs over \cite{dou2023sensing} and \cite{liu2023toward}, respectively. Furthermore, Fig. \ref{two}(c) and Fig. \ref{two}(d) demonstrate that how HDVs can affect the performance of the system. Similarly, more HDVs result in fewer TTCs and higher CRBs. Compared to \cite{dou2023sensing} and \cite{liu2023toward} respectively, our proposed model increases TTCs by 33\% and 53\%, as well as decreases CRBs by 72\% and 83\%.
%In the case of more RSUs, every RSU is responsible for fewer CAVs. Furthermore, the performance of the system increases and we see an upward trend for TTC in Fig. \ref{two}(e) and a downward trend for CRBs in Fig. \ref{two}(f). Our model outperforms \cite{dou2023sensing} and \cite{liu2023toward} by 24\% and 43\% improvement in TTCs, as well as 7\% and 8\% reduction in CRBs, respectively. Similarly, with increase in the number of antennas of each RSU, we see the same trends in Fig. \ref{three}(a) and \ref{three}(b). This can be attributed to the fact that deploying large-scale antenna arrays can bring an improved antenna gain to the RSU system. Our model improves TTCs by 26\% and 42\% as well as reduces CRBs by 6\% and 7\% compared to \cite{dou2023sensing} and \cite{liu2023toward}, respectively.
When the number of RSUs increases, each RSU serves fewer CAVs, thereby enhancing overall system performance. As shown in Fig. \ref{two}(e) and Fig. \ref{two}(f), this leads to an upward trend in TTC and a downward trend in CRBs. Compared to \cite{dou2023sensing} and \cite{liu2023toward}, our model achieves 24\% and 43\% improvement in TTC, along with 7\% and 8\% reduction in CRBs, respectively. A similar trend is observed when increasing the number of antennas per RSU, as depicted in Fig. \ref{three}(a) and Fig. \ref{three}(b). This improvement is attributed to the enhanced antenna gain provided by large-scale antenna arrays. In this case, our model achieves 26\% and 42\% improvement in TTC and reduces CRBs by 6\% and 7\% relative to \cite{dou2023sensing} and \cite{liu2023toward}, respectively.

\begin{figure*}[!t]
	\centering
	\subfigure[]{\includegraphics[width=0.3\textwidth]{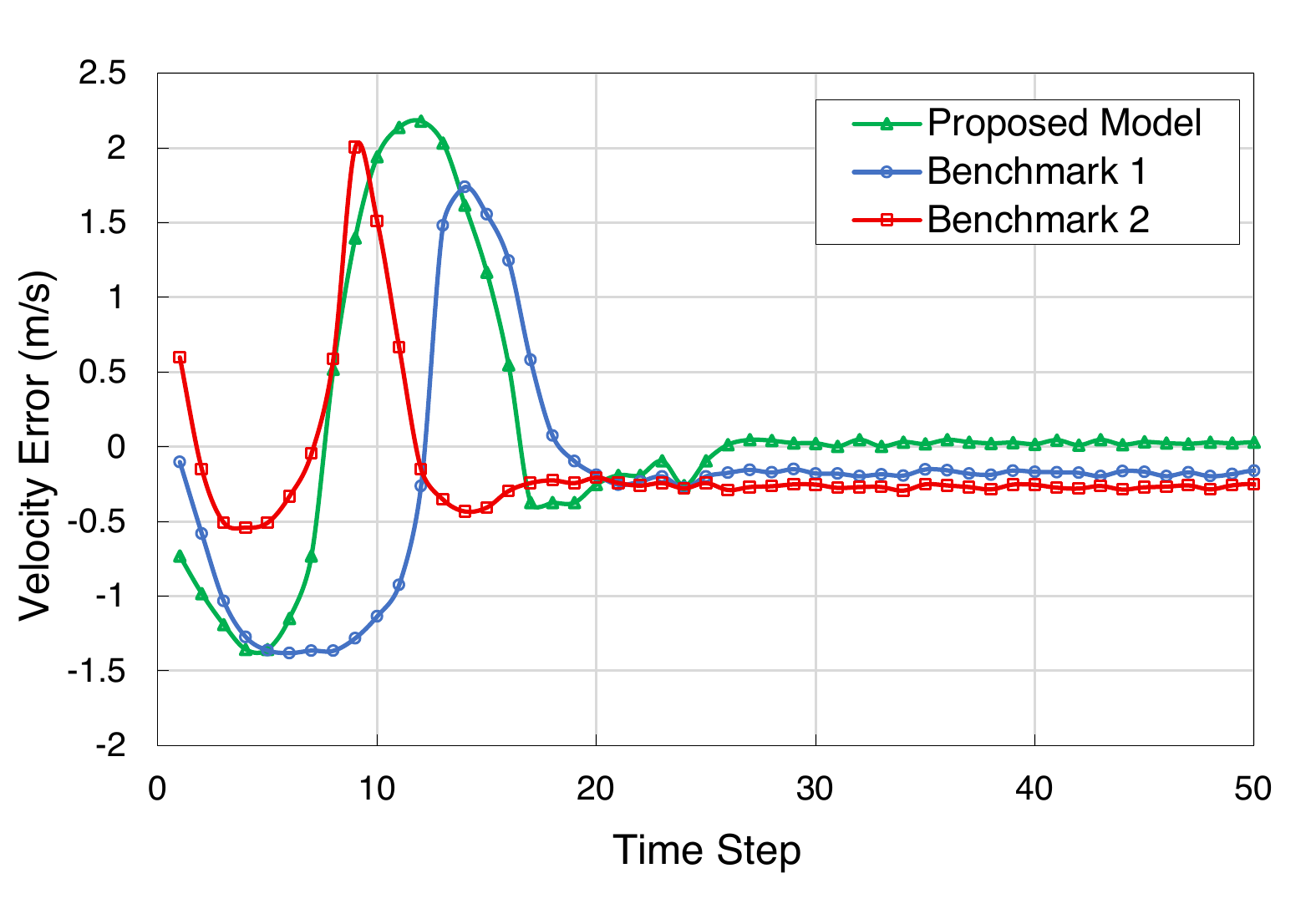}}
	\subfigure[]{\includegraphics[width=0.3\textwidth]{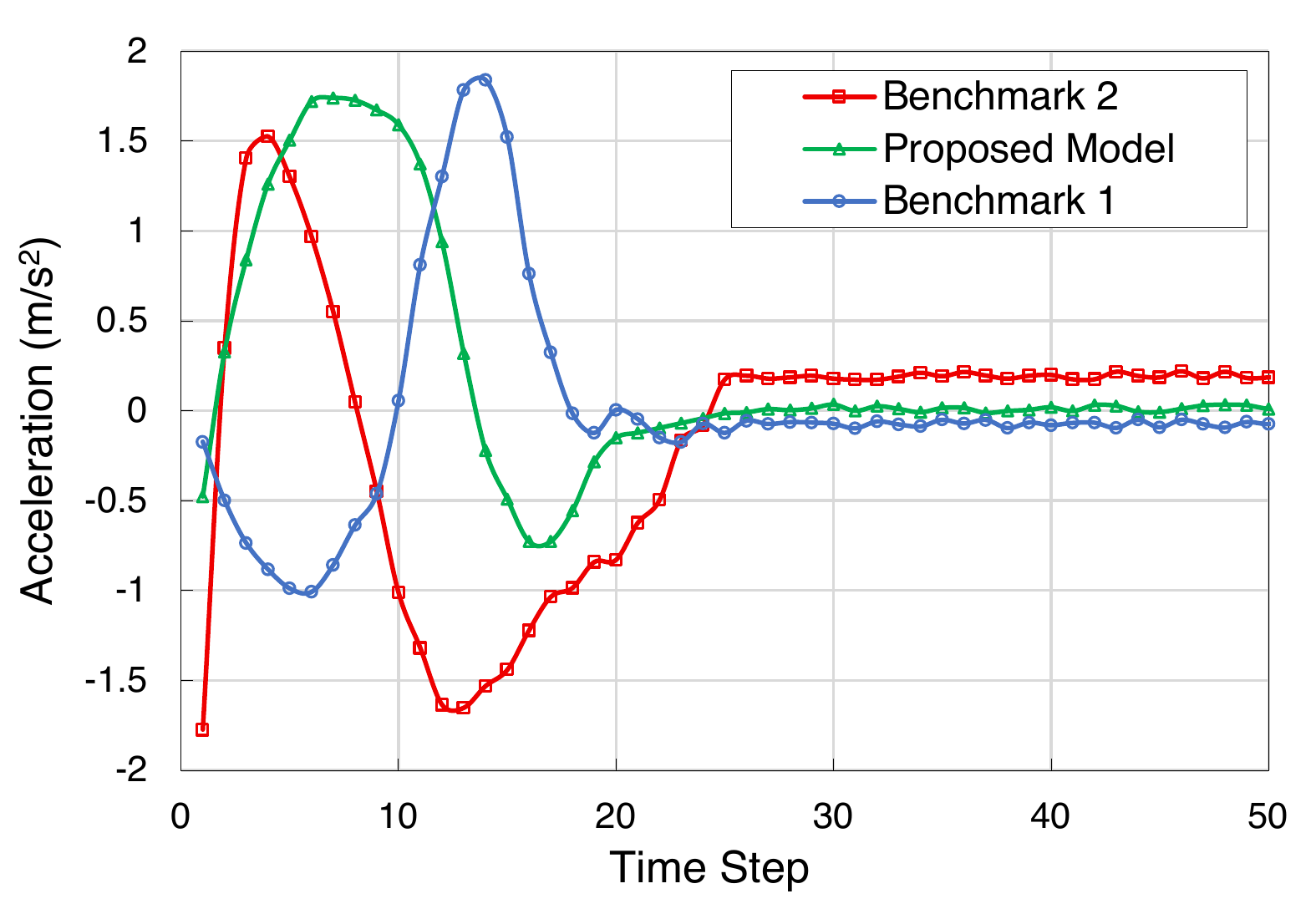}}
	\subfigure[]{\includegraphics[width=0.3\textwidth]{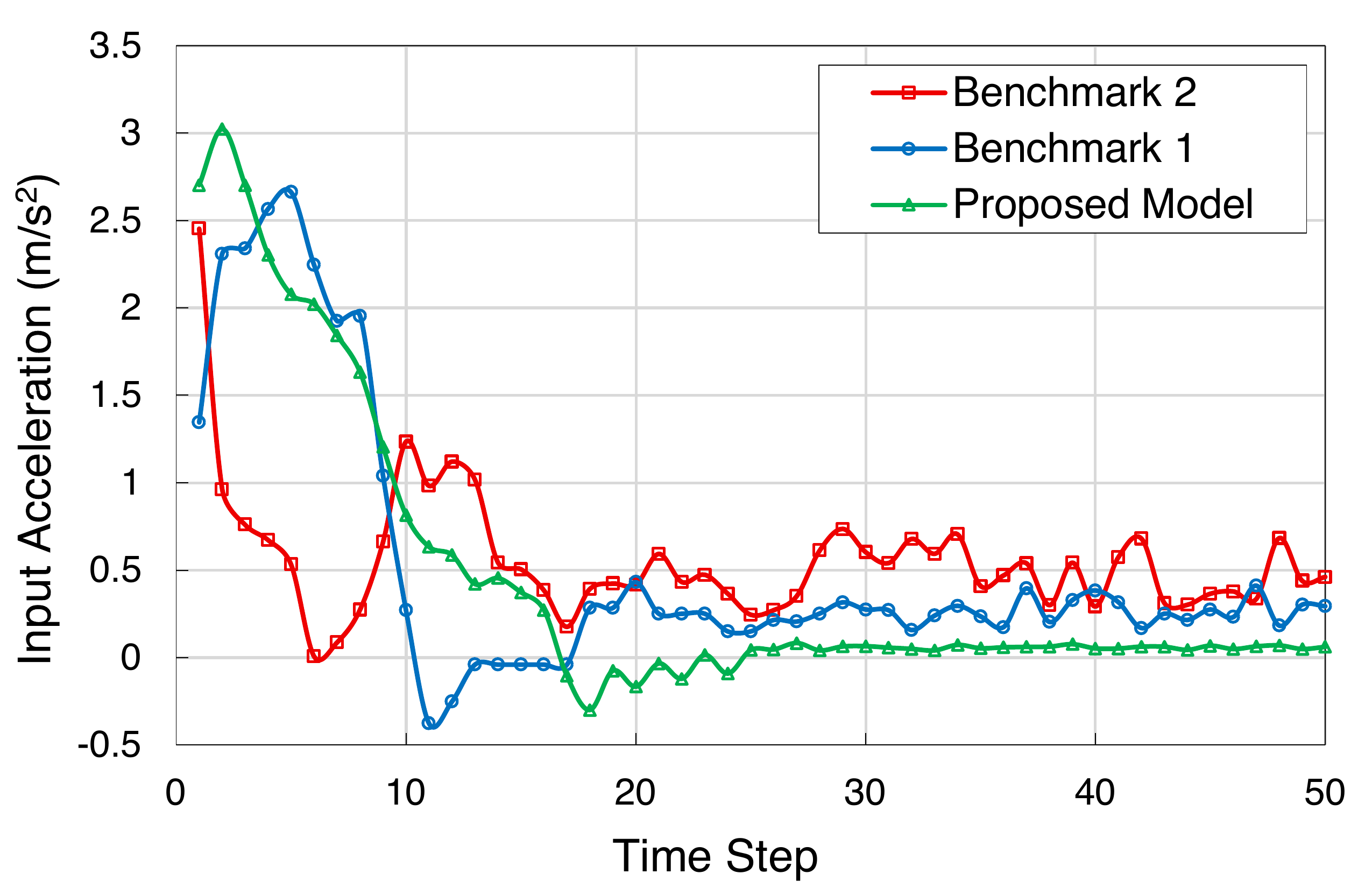}}
	\caption{(a) velocity error, (b) acceleration, and (c) input acceleration over time}
	\label{four}
\end{figure*}

%Besides, Fig. \ref{three}(c) and Fig. \ref{three}(d) illustrate our model performance for different transmit powers. As shown in Fig. \ref{three}(c), for more transmit power budgets, TTCs increases because a higher power improves the received SNR at the RSU and thus more vehicles’ information can be sent. Additionally, with increase in the transmit powers, CRBs decreases in Fig. \ref{three}(d). This is because that similarly, a higher  power improves the received SNR, resulting in a more accurate estimation for RSUs. Our proposed model demonstrates 54\% and 80\% improvement in TTCs as well as 7\% and 8\% reduction in CRBs relative to \cite{dou2023sensing} and \cite{liu2023toward}, respectively.
Furthermore, Fig. \ref{three}(c) and Fig. \ref{three}(d) present the performance of our model under varying transmit power budgets. As observed in Fig. \ref{three}(c), higher transmit power leads to an increase in TTC, since the improved received SNR at the RSU enables transmission of more vehicles’ information. Similarly, Fig. \ref{three}(d) shows that CRBs decrease with increasing transmit power, as the enhanced SNR yields more accurate RSU estimations. Overall, our proposed model achieves up to 54\% and 80\% improvement in TTC, along with 7\% and 8\% reduction in CRBs, compared to \cite{dou2023sensing} and \cite{liu2023toward}, respectively.
%\subsection{KL Estimation}
%Fig. \ref{three}(e) shows the KL divergence for each time step. It can be seen that it starts at a relatively high value near 0.9 and then decreases gradually over time. So transmitting the corresponding V2X information and including them in the state space can help better predict the next state.
%%\subsection{Driving Status and Control Inputs}
%Fig. \ref{three}(f) and Fig. \ref{four} show the spacing and velocity error as well as accelaretion for the proposed and baseline models. At the beginning of the episode, the input acceleration $u^l_{v,\tau}$ remains the maximum value to increase the acceleration $z^{l}_{v,t_\tau}$ as promptly as possible, so that the control errors $e^{l}_{v,t_\tau}$ and $\tilde{e}^{l}_{v,t_\tau}$ can be promptly reduced. We can see in Fig. \ref{four}(c) that the input acceleration $u^l_{v,\tau}$ for \cite{dou2023sensing} and \cite{liu2023toward} exhibits higher fluctuation than those of the proposed model, which means that the vehicle supported by our model drives more smoothly.
Fig. \ref{three}(e) illustrates the KL divergence across time steps. It starts at a relatively high value near 0.9 and gradually decreases, indicating that transmitting the corresponding V2X information and incorporating it into the state space enhances next-state prediction. Fig. \ref{three}(f) and Fig. \ref{four} present the spacing error, velocity error, and acceleration for both the proposed and baseline models. At the beginning of each episode, the input acceleration $u^l_{v,\tau}$ is maintained at its maximum to rapidly increase $z^{l}_{v,t_\tau}$, thereby promptly reducing the control errors $e^{l}_{v,t_\tau}$ and $\tilde{e}^{l}_{v,t_\tau}$. As shown in Fig. \ref{four}(c), the input acceleration $u^l_{v,\tau}$ in \cite{dou2023sensing} and \cite{liu2023toward} exhibits larger fluctuations compared to our proposed model, indicating that vehicles supported by our framework achieve smoother driving dynamics.

\section{Summary and Conclusion}\label{Conclusion}
%In this paper, we presented a novel scheme for enhancing vehicular safety and positioning accuracy in 6G-enabled V2X networks by integrating sensing and communication capabilities through the VoI paradigm. By introducing a safety-aware ISAC-assisted system, we focused on prioritizing and transmitting only the most safety-critical information, thereby alleviating the limitations of bandwidth and latency inherent in ultra-dense vehicular environments. We formulated the sensing, communication, and control tasks as a two-time-scale sequential stochastic decision problem and proposed a multi-agent deep reinforcement learning-based solution using a MA-DDPG algorithm. This approach effectively tackled the curse of dimensionality by constructing a compact and high-impact state representation, enabling efficient control of CAVs while maximizing safety metrics such as TTC and minimizing the CR ratio. Simulation results demonstrated that our VoI-driven method significantly increases the TTC across all CAVs compared to the benchmarks up to 66\%. The KL divergence analysis further validated the system's ability to discern and select high-value exogenous information, contributing to robust decision-making and control in dynamic mixed-traffic scenarios. Overall, the proposed framework lays a foundation for future 6G ITS systems that demand precise positioning, reliable communication, and real-time safety assurance.
This paper introduces a novel VoI-driven framework for enhancing vehicular safety and positioning accuracy in 6G-enabled V2X networks through ISAC. A safety-aware mechanism prioritizes transmission of only safety-critical information, mitigating bandwidth and latency constraints in ultra-dense traffic. The sensing, communication, and control tasks are modeled as a two-time-scale sequential stochastic decision problem, solved via a MA-DDPG-based multi-agent reinforcement learning approach. By constructing a compact, high-impact state representation, the method addresses the curse of dimensionality and enables efficient control of CAVs. Simulation results show significant safety gains, with TTC improved by up to 66\% and CR minimized compared to benchmarks. KL divergence analysis confirms effective selection of high-value exogenous information, ensuring robust decision-making in dynamic mixed-traffic environments. The proposed framework establishes a foundation for future 6G ITS systems requiring precise positioning, reliable communication, and real-time safety assurance.

\hyphenation{op-tical net-works semi-conduc-tor}

\ifCLASSOPTIONcaptionsoff
\newpage
\fi

\bibliographystyle{IEEEtran}
\bibliography{Bibliography}

% Generated by IEEEtran.bst, version: 1.14 (2015/08/26)
\begin{thebibliography}{10}
\providecommand{\url}[1]{#1}
\csname url@samestyle\endcsname
\providecommand{\newblock}{\relax}
\providecommand{\bibinfo}[2]{#2}
\providecommand{\BIBentrySTDinterwordspacing}{\spaceskip=0pt\relax}
\providecommand{\BIBentryALTinterwordstretchfactor}{4}
\providecommand{\BIBentryALTinterwordspacing}{\spaceskip=\fontdimen2\font plus
\BIBentryALTinterwordstretchfactor\fontdimen3\font minus
  \fontdimen4\font\relax}
\providecommand{\BIBforeignlanguage}[2]{{%
\expandafter\ifx\csname l@#1\endcsname\relax
\typeout{** WARNING: IEEEtran.bst: No hyphenation pattern has been}%
\typeout{** loaded for the language `#1'. Using the pattern for}%
\typeout{** the default language instead.}%
\else
\language=\csname l@#1\endcsname
\fi
#2}}
\providecommand{\BIBdecl}{\relax}
\BIBdecl

\bibitem{10639496}
K.~Hou and S.~Zhang, ``Optimal beamforming for secure integrated sensing and
  communication exploiting target location distribution,'' \emph{IEEE Journal
  on Selected Areas in Communications}, vol.~42, no.~11, pp. 3125--3139,
  November 2024.

\bibitem{10845207}
L.~Li, J.~Zhang, and T.-H. Chang, ``Beamforming optimization for robust sensing
  and communication in dynamic mmwave {MIMO} networks,'' \emph{IEEE Journal on
  Selected Areas in Communications}, vol.~43, no.~4, pp. 1354--1370, April
  2025.

\bibitem{10556618}
X.~Gan, C.~Huang, Z.~Yang, X.~Chen, J.~He, Z.~Zhang, C.~Yuen, Y.~Liang~Guan,
  and M.~Debbah, ``Coverage and rate analysis for integrated sensing and
  communication networks,'' \emph{IEEE Journal on Selected Areas in
  Communications}, vol.~42, no.~9, pp. 2213--2227, September 2024.

\bibitem{9791349}
C.~Liu, W.~Yuan, S.~Li, X.~Liu, H.~Li, D.~W.~K. Ng, and Y.~Li, ``Learning-based
  predictive beamforming for integrated sensing and communication in vehicular
  networks,'' \emph{IEEE Journal on Selected Areas in Communications}, vol.~40,
  no.~8, pp. 2317--2334, August 2022.

\bibitem{9728752}
Q.~Zhang, H.~Sun, X.~Gao, X.~Wang, and Z.~Feng, ``Time-division {ISAC} enabled
  connected automated vehicles cooperation algorithm design and performance
  evaluation,'' \emph{IEEE Journal on Selected Areas in Communications},
  vol.~40, no.~7, pp. 2206--2218, July 2022.

\bibitem{9724187}
Z.~Xiao and Y.~Zeng, ``Waveform design and performance analysis for full-duplex
  integrated sensing and communication,'' \emph{IEEE Journal on Selected Areas
  in Communications}, vol.~40, no.~6, pp. 1823--1837, June 2022.

\bibitem{10960374}
S.~Xu, H.~Sun, Y.~Xu, T.~Guo, C.~Li, and L.~Yang, ``Distributed compression
  method for channel calibration in cell-free {MIMO} {ISAC} systems,''
  \emph{IEEE Journal on Selected Areas in Communications}, vol.~43, no.~7, pp.
  2349--2363, July 2025.

\bibitem{10032141}
P.~Gao, L.~Lian, and J.~Yu, ``Cooperative {ISAC} with direct localization and
  rate-splitting multiple access communication: A pareto optimization
  framework,'' \emph{IEEE Journal on Selected Areas in Communications},
  vol.~41, no.~5, pp. 1496--1515, May 2023.

\bibitem{10158711}
Z.~He, W.~Xu, H.~Shen, D.~W.~K. Ng, Y.~C. Eldar, and X.~You, ``Full-duplex
  communication for {ISAC}: Joint beamforming and power optimization,''
  \emph{IEEE Journal on Selected Areas in Communications}, vol.~41, no.~9, pp.
  2920--2936, September 2023.

\bibitem{9171304}
F.~Liu, W.~Yuan, C.~Masouros, and J.~Yuan, ``Radar-assisted predictive
  beamforming for vehicular links: Communication served by sensing,''
  \emph{IEEE Transactions on Wireless Communications}, vol.~19, no.~11, pp.
  7704--7719, November 2020.

\bibitem{9246715}
W.~Yuan, F.~Liu, C.~Masouros, J.~Yuan, D.~W.~K. Ng, and N.~González-Prelcic,
  ``Bayesian predictive beamforming for vehicular networks: A low-overhead
  joint radar-communication approach,'' \emph{IEEE Transactions on Wireless
  Communications}, vol.~20, no.~3, pp. 1442--1456, March 2021.

\bibitem{9724174}
L.~Chen, Z.~Wang, Y.~Du, Y.~Chen, and F.~R. Yu, ``Generalized transceiver
  beamforming for {DFRC} with {MIMO} radar and {MU-MIMO} communication,''
  \emph{IEEE Journal on Selected Areas in Communications}, vol.~40, no.~6, pp.
  1795--1808, June 2022.

\bibitem{10061429}
X.~Meng, F.~Liu, C.~Masouros, W.~Yuan, Q.~Zhang, and Z.~Feng, ``Vehicular
  connectivity on complex trajectories: Roadway-geometry aware {ISAC}
  beam-tracking,'' \emph{IEEE Transactions on Wireless Communications},
  vol.~22, no.~11, pp. 7408--7423, November 2023.

\bibitem{9229189}
P.~Yang, D.~Duan, C.~Chen, X.~Cheng, and L.~Yang, ``Multi-sensor multi-vehicle
  ({MSMV}) localization and mobility tracking for autonomous driving,''
  \emph{IEEE Transactions on Vehicular Technology}, vol.~69, no.~12, pp.
  14\,355--14\,364, December 2020.

\bibitem{9800700}
W.~Jiang, Z.~Wei, B.~Li, Z.~Feng, and Z.~Fang, ``Improve radar sensing
  performance of multiple roadside units cooperation via space registration,''
  \emph{IEEE Transactions on Vehicular Technology}, vol.~71, no.~10, pp.
  10\,975--10\,990, October 2022.

\bibitem{10504613}
Y.~He, H.~Zhao, and S.~Shao, ``Nonlinear self-interference cancellation in
  vehicle networks for full-duplex integrated sensing and communication,''
  \emph{IEEE Transactions on Vehicular Technology}, vol.~73, no.~9, pp.
  13\,980--13\,985, September 2024.

\bibitem{9820762}
Z.~Wang, K.~Han, J.~Jiang, F.~Liu, and W.~Yuan, ``Multi-vehicle tracking and
  {ID} association based on integrated sensing and communication signaling,''
  \emph{IEEE Wireless Communications Letters}, vol.~11, no.~9, pp. 1960--1964,
  September 2022.

\bibitem{10677488}
J.~Singh, A.~Gupta, A.~K. Jagannatham, and L.~Hanzo, ``Multi-beam
  object-localization for millimeter-wave {ISAC-Aided} connected autonomous
  vehicles,'' \emph{IEEE Transactions on Vehicular Technology}, vol.~74, no.~1,
  pp. 1725--1729, January 2025.

\bibitem{10433790}
Z.~Ye, C.~Yu, H.~Zhu, Y.~He, M.~Gao, and G.~Yu, ``{ISAC}-assisted collision
  avoidance mechanism for vehicle-to-infrastructure systems,'' \emph{IEEE
  Transactions on Intelligent Vehicles}, vol.~9, no.~10, pp. 6242--6257,
  October 2024.

\bibitem{10502156}
Y.~Li, F.~Liu, Z.~Du, W.~Yuan, Q.~Shi, and C.~Masouros, ``Frame structure and
  protocol design for sensing-assisted {NR-V2X} communications,'' \emph{IEEE
  Transactions on Mobile Computing}, vol.~23, no.~12, pp. 11\,045--11\,060,
  December 2024.

\bibitem{10738493}
Z.~Xu, S.~Xu, H.~Ding, and R.~Xu, ``An {ISAC}-based beam tracking scheme
  against inter-region interference for the multi-{RSU} {V2I} scenario,''
  \emph{IEEE Transactions on Vehicular Technology}, vol.~74, no.~3, pp.
  4257--4272, March 2025.

\bibitem{9947033}
Z.~Du, F.~Liu, W.~Yuan, C.~Masouros, Z.~Zhang, S.~Xia, and G.~Caire,
  ``Integrated sensing and communications for {V2I} networks: Dynamic
  predictive beamforming for extended vehicle targets,'' \emph{IEEE
  Transactions on Wireless Communications}, vol.~22, no.~6, pp. 3612--3627,
  June 2023.

\bibitem{1-choudhury2020experimental}
B.~Choudhury, V.~K. Shah, A.~Dayal, and J.~H. Reed, ``Experimental analysis of
  safety application reliability in {V2V} networks,'' in \emph{Proceedings of
  IEEE 91st Vehicular Technology Conference (VTC2020-Spring)}, Antwerp,
  Belgium, May 2020, pp. 1--5.

\bibitem{aznar2019time}
J.~Aznar-Poveda, E.~Egea-Lopez, A.-J. Garcia-Sanchez, and P.~Pavon-Mari{\'a}o,
  ``Time-to-collision-based awareness and congestion control for vehicular
  communications,'' \emph{IEEE Access}, vol.~7, pp. 154\,192--154\,208, 2019.

\bibitem{ETSITR1033001}
{ETSI}, ``{ETSI TR 103.300‑1: Integrated Satellite Communications; Interfaces
  between mobile‑satellite services and terrestrial networks (Part 1)},''
  {ETSI}, {Technical Report} TR 103.300‑1, 2024.

\bibitem{ETSITS1015393}
------, ``{ETSI TS 101.539‑3: Digital Cellular Telecommunications System
  (Phase 2+); Universal Mobile Telecommunications System (UMTS);
  Mission‑Critical Services; MC‑SUBUIL‑T Architecture (Part 3)},''
  {ETSI}, {Standard} TS 101.539‑3, 2023.

\bibitem{11020763}
A.~S. Sümer, E.~Memişoğlu, and H.~Arslan, ``A novel pilot allocation
  technique for uplink {OFDMA} in {ISAC} systems,'' \emph{IEEE Wireless
  Communications Letters}, pp. 1--1, Early Access 2025.

\bibitem{yetis2021joint}
C.~M. Yetis, E.~Bj{\"o}rnson, and P.~Giselsson, ``Joint analog beam selection
  and digital beamforming in millimeter wave cell-free massive {MIMO}
  systems,'' \emph{IEEE Open Journal of the Communications Society}, vol.~2,
  pp. 1647--1662, 2021.

\bibitem{10804675-1}
Z.~Chen and Q.~Liang, ``Massive {MIMO} channel modeling based on {B5G/6G} and
  consumer {IoT},'' \emph{IEEE Transactions on Consumer Electronics}, pp. 1--1,
  Early Access 2024.

\bibitem{10571114}
Y.~Cui, F.~Liu, W.~Yuan, J.~Mu, X.~Jing, and D.~W. Kwan~Ng, ``Optimal precoding
  design for monostatic {ISAC} systems: {MSE} lower bound and {DoF}
  completion,'' in \emph{Proccedings of IEEE Wireless Communications and
  Networking Conference (WCNC)}, Dubai, United Arab Emirates, April 2024, pp.
  1--6.

\bibitem{10736521}
B.~M. Lee, ``Efficient resource management for massive {MIMO} in high-density
  massive {IoT} networks,'' \emph{IEEE Transactions on Mobile Computing},
  vol.~24, no.~3, pp. 1963--1980, March 2025.

\bibitem{10484981}
H.~Fang, X.~Qiao, Y.~Zhang, L.~Yang, and H.~Zhu, ``On the performance of
  {RIS}-aided cell-free massive {MIMO} systems under channel aging,''
  \emph{IEEE Transactions on Vehicular Technology}, vol.~73, no.~9, pp.
  12\,828--12\,841, 2024.

\bibitem{8770141}
W.~Yuan, N.~Wu, Q.~Guo, X.~Huang, Y.~Li, and L.~Hanzo, ``{TOA}-based passive
  localization constructed over factor graphs: A unified framework,''
  \emph{IEEE Transactions on Communications}, vol.~67, no.~10, pp. 6952--6965,
  October 2019.

\bibitem{10745726}
Z.~Yuan, L.~Yu, Z.~Wang, C.~Li, T.~Dallmann, and W.~Fan, ``Experimental
  analysis and modeling of monostatic {AAV} {RCS} for {ISAC} channels,''
  \emph{IEEE Antennas and Wireless Propagation Letters}, vol.~24, no.~1, pp.
  222--226, January 2025.

\bibitem{Qiu2019}
C.~Qiu, Y.~Hu, Y.~Chen, and B.~Zeng, ``Deep deterministic policy gradient
  ({DDPG})-based energy harvesting wireless communications,'' \emph{IEEE
  Internet of Things Journal}, vol.~6, no.~5, pp. 8577--8588, 2019.

\bibitem{vu2020cell}
Y.~Cao, Z.~Zhang, X.~Xia, P.~Xin, D.~Liu, K.~Zheng, M.~Lou, J.~Jin, Q.~Wang,
  D.~Wang, Y.~Huang, X.~You, and J.~Wang, ``Implementation of a cell-free {RAN}
  system with distributed cooperative transceivers under {ORAN} architecture,''
  \emph{IEEE Journal on Selected Areas in Communications}, vol.~43, no.~3, pp.
  765--779, March 2025.

\bibitem{dou2023sensing}
C.~Dou, N.~Huang, Y.~Wu, L.~Qian, and T.~Q. Quek, ``Sensing-efficient
  {NOMA}-aided integrated sensing and communication: A joint sensing scheduling
  and beamforming optimization,'' \emph{IEEE Transactions on Vehicular
  Technology}, vol.~72, no.~10, pp. 13\,591--13\,603, 2023.

\bibitem{liu2023toward}
Z.~Liu, X.~Li, H.~Ji, H.~Zhang, and V.~C. Leung, ``Toward {STAR-RIS}-empowered
  integrated sensing and communications: Joint active and passive beamforming
  design,'' \emph{IEEE Transactions on Vehicular Technology}, vol.~72, no.~12,
  pp. 15\,991--16\,005, 2023.

\end{thebibliography}
\end{document}